\documentclass[sigconf, 9pt]{acmart}

\AtBeginDocument{%
  }

\makeatletter
\def\@ACM@copyright@check@cc{}
\makeatother

\acmYear{2026}\copyrightyear{2026}
\setcopyright{cc}
\setcctype{by}
\acmConference[Middleware '26]{27th International Middleware Conference}{December 14--18, 2026}{Tarragona, Spain}
\acmBooktitle{27th International Middleware Conference (Middleware '26), December 14--18, 2026, Tarragona, Spain}
\acmDOI{10.1145/3801927.3810468}
\acmISBN{979-8-4007-2621-7/26/11}

\usepackage{makecell}
\usepackage{graphicx}
\usepackage{flushend}
\usepackage{caption}
\usepackage{subfig}
\usepackage{comment}
\usepackage{algorithm}
\usepackage{algpseudocode}
\usepackage[utf8]{inputenc}
\usepackage{kotex}
\usepackage{multirow}
\usepackage{array}
\usepackage{tabularx}
\usepackage{booktabs}
\usepackage{pifont}
\usepackage{xstring}
\usepackage{enumitem}
\usepackage{microtype}
\usepackage{color,soul}
\usepackage{colortbl}
\usepackage{xcolor}

\definecolor{hlcolor}{RGB}{255, 255, 0}

\newcolumntype{Y}{>{\raggedright\arraybackslash}X}

\let\oldtexttt\texttt
\renewcommand{\texttt}[1]{\oldtexttt{\StrSubstitute[0]{#1}{.}{.\allowbreak}}}

\begin{document}

\title[KubePACS: \\\textbf{Kube}rnetes Cluster Using \textbf{P}erformant, Highly \textbf{A}vailable, and \textbf{C}ost Efficient \textbf{S}pot Instances]{KubePACS: \textsl{Kube}rnetes Cluster Using \textsl{P}erformant, Highly \textsl{A}vailable, and \textsl{C}ost Efficient \textsl{S}pot Instances}

\settopmatter{authorsperrow=3}

\author{Taeyoon Kim}
\email{tykim7@hanyang.ac.kr}
\orcid{0009-0003-0705-0713}
\affiliation{%
  \department{Dept. of Data Science}
  \institution{Hanyang University}
  \city{Seoul}
  \country{Republic of Korea}
}
\authornote{Both authors contributed equally to this work.}
\author{Kyumin Kim}
\email{okkimok123@hanyang.ac.kr}
\orcid{0009-0007-2705-1122}
\affiliation{%
  \department{Dept. of Data Science}
  \institution{Hanyang University}
  \city{Seoul}
  \country{Republic of Korea}
}
\authornotemark[1]

\author{Enrique Molina-Giménez}
\email{enrique.molina@urv.cat}
\orcid{0009-0005-2597-3815}
\affiliation{%
  \department{Dept. of Computer Eng. and Math.}
  \institution{Universitat Rovira i Virgili}
  \city{Tarragona}
  \country{Spain}
}
\author{Pedro García-López}
\email{pedro.garcia@urv.cat}
\orcid{0000-0002-9848-1492}
\affiliation{%
  \department{Dept. of Computer Eng. and Math.}
  \institution{Universitat Rovira i Virgili}
  \city{Tarragona}
  \country{Spain}
}

\author{Kyungyong Lee}
\email{kyungyong@hanyang.ac.kr}
\orcid{0000-0003-0312-4386}
\affiliation{%
  \department{Dept. of Data Science}
  \institution{Hanyang University}
  \city{Seoul}
  \country{Republic of Korea}
}
\authornote{Corresponding author}

\begin{abstract}
  Cloud users aim to minimize cost while maximizing performance by selecting the most suitable instance types for their workloads. To reduce expenses, spot instances have been widely adopted due to their steep discounts compared to on-demand pricing. However, their use introduces reliability risks due to potential interruptions, and existing research has primarily focused on mitigating this trade-off from a cost or availability perspective alone. Despite the diversity in hardware capabilities among instance types, current provisioning systems tend to ignore performance variation, selecting nodes solely based on minimum resource requirements.

  In this paper, we present KubePACS, a Kubernetes-native spot instance provisioning system that constructs node pools optimized for both cost and performance while guaranteeing high availability. KubePACS formulates the node selection process as a multi-objective optimization problem, incorporating real-time data such as spot prices, performance benchmarks, and availability scores, including the multi-node Spot Placement Score (SPS). It solves this problem efficiently using an Integer Linear Programming (ILP) approach guided by the Golden Section Search (GSS) algorithm to find the optimal configuration. By integrating with the Karpenter node autoscaler, KubePACS jointly optimizes instance-type selection and node scaling decisions within a standard provisioning workflow. KubePACS also adopts a novel heuristic to support workload-specific preferences by scaling performance metrics for specialized instances. Through extensive evaluation across synthetic and real-world workloads, KubePACS demonstrates on average 55.09\% and up to 81.06\% higher performance per dollar over state-of-the-art solutions such as Karpenter, SpotVerse, and SpotKube, which only reference the spot instance prices and limited availability data.
\end{abstract}


\maketitle

\section{Introduction}
\begin{figure*}[ht!]
  \includegraphics[width=0.7\textwidth]{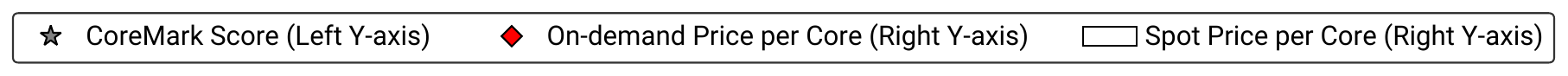}

  \vspace{-1.2em}
  \centering
  \subfloat[Intel general optimized series]{
    \includegraphics[width=0.242 \textwidth]{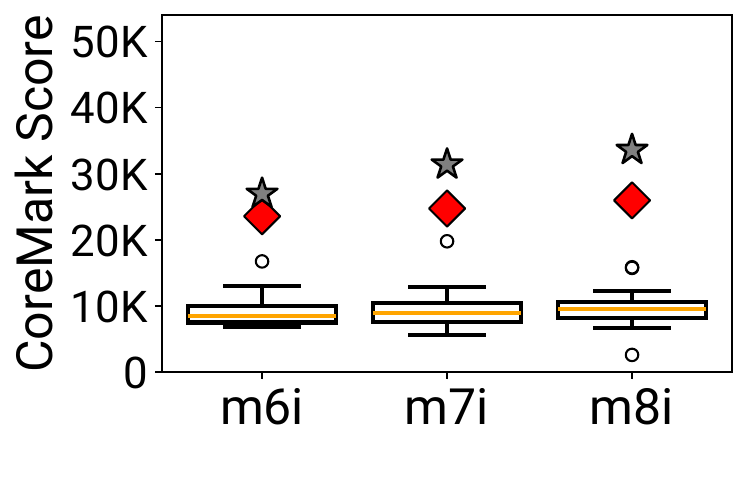}
    \label{fig:instance-generation-coremark-price-variation}
  }
  \hspace{-0.5em}
  \subfloat[7th generation instance series]{
    \includegraphics[width=0.242 \textwidth]{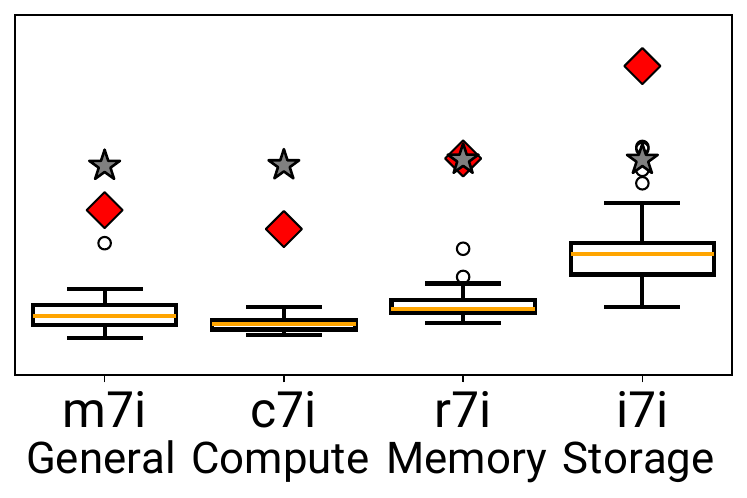}
    \label{fig:instance-series-coremark-price-variation}
  }
  \hspace{-0.4em}
  \subfloat[M6i family options]{
    \includegraphics[width=0.242 \textwidth]{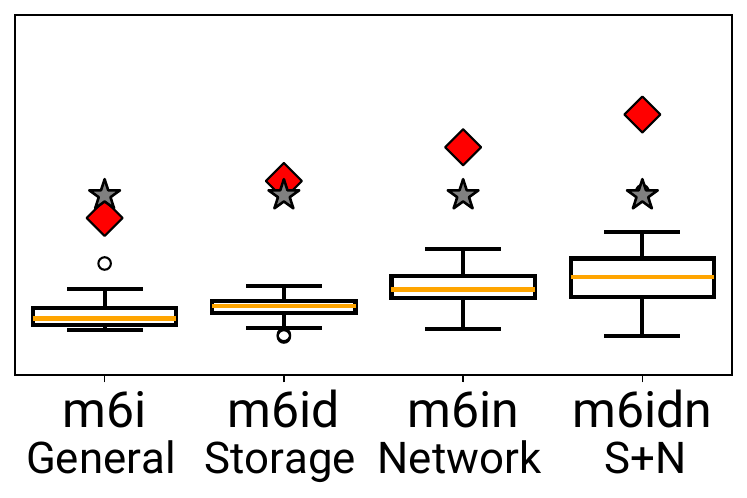}
    \label{fig:instance-options-coremark-price-variation}
  }
  \hspace{-0.4em}
  \subfloat[CPU Vendor]{
    \includegraphics[width=0.242 \textwidth]{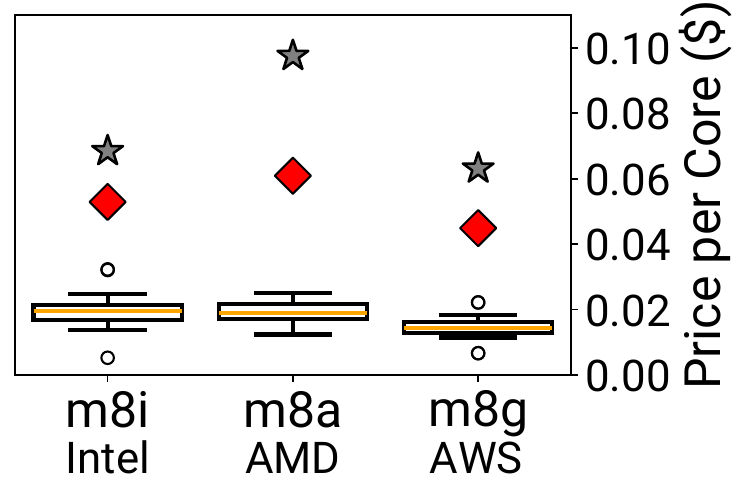}
    \label{fig:vendor-comparison-coremark-price}
  }
  \caption{Comparing benchmark score (CoreMark) and spot instance price variation. Different instance configurations show significant variations for on-demand price, hardware performance, and spot prices}
  \Description{Comparing benchmark score (CoreMark) and spot instance price variation. Different instance configurations show significant variations for on-demand price, hardware performance, and spot prices}
  \label{fig:coremark-and-price-variation}
\end{figure*}

Cloud computing has evolved into a paradigm that prioritizes performance optimization and cost-efficiency through fine-grained control over infrastructure composition. Modern cloud providers offer a wide variety of instance types with diverse compute capabilities, storage I/O throughput, and network bandwidth, enabling users to tailor infrastructure to workload-specific performance demands. At the same time, dynamic pricing models, such as spot instances, further incentivize cost savings by allowing users to access unused capacity at substantial discounts.

However, selecting optimal instance types is non-trivial. Users are often forced to choose instances based solely on static resource requirements, such as CPU cores and memory size, without fully accounting for differences in hardware performance across instance families. This leads to missed opportunities for performance-per-dollar optimization, especially in large-scale systems.

While spot instances offer compelling cost benefits of up to 90\% cheaper than on-demand instances, their volatility due to potential interruptions presents additional challenges for reliable cluster formation. Although real-time availability metrics, such as Spot Placement Score (SPS) offered by Amazon Web Services (AWS)~\cite{spot-placement-score-start} and Microsoft Azure~\cite{azure-spot-placement-score}, have been introduced by cloud providers to improve visibility into reliability, current provisioning tools underutilize this information and lack the ability to integrate it with performance and price data in a unified optimization process.

Kubernetes, as the de facto standard for container orchestration, demands intelligent node provisioning mechanisms considering application performance, cost-efficiency, and reliability when using spot instances. Existing tools such as Karpenter~\cite{karpenter} focus primarily on satisfying resource constraints, such as the total number of pods or per-pod CPU and memory requirements, without accounting for differences in hardware performance, cost-performance tradeoffs, or the availability dynamics of multiple spot instances.

SpotKube~\cite{spotkube}, a Kubernetes node provisioning framework utilizing spot instances, aims to reduce costs by applying a genetic algorithm. However, it considers only spot instance prices and disregards performance heterogeneity and availability indicators, limiting its effectiveness in maximizing performance-per-dollar or ensuring robustness under spot volatility. SpotVerse~\cite{spotverse} improves on availability awareness by incorporating datasets such as SPS and Interruption Frequency (IF). Nonetheless, it infers large-scale availability based solely on single-node SPS metrics, which are known to be imprecise for multi-node provisioning~\cite{multinode-spot-dataset}. Moreover, it does not incorporate hardware performance benchmarks into its decision-making, which results in potentially suboptimal instance selection in terms of computational efficiency.

In this paper, we present KubePACS, a Kubernetes-native instance provisioning system that overcomes the limitations of prior approaches by integrating real-time, multi-dimensional spot instance datasets into the node selection process. KubePACS formulates node provisioning as a multi-objective optimization problem that jointly considers spot price, hardware performance, and large-scale availability metrics. To the best of our knowledge, KubePACS is the first system to jointly utilize multi-node SPS data, hardware performance benchmarks, and spot pricing within a unified optimization framework for cluster-level provisioning. To support workloads with specific I/O characteristics, such as network- or disk-intensive applications, KubePACS applies a workload-aware scaling heuristic that adjusts performance scores by leveraging on-demand price to reflect instance specialization. The system employs an ILP solver guided by a tunable cost-performance weight, and efficiently searches for the optimal trade-off using Golden Section Search (GSS)~\cite{golden-section-search}. By integrating directly into the Kubernetes node autoscaler at the controller level, KubePACS bridges the gap between research prototype and production-ready deployment.

We evaluate KubePACS on a variety of synthetic and real-world workloads, benchmarking its effectiveness against state-of-the-art provisioning systems, including SpotVerse~\cite{spotverse}, SpotKube~\cite{spotkube}, and Karpenter~\cite{karpenter}, within the AWS cloud environment. In large-scale Kubernetes cluster provisioning scenarios, KubePACS achieves a 81.06\% improvement in cost-performance efficiency compared to SpotVerse, while also enhancing the robustness of multi-node spot instance availability. This gain is attributed to KubePACS's integration of critical datasets, including spot pricing, hardware performance benchmarks, and multi-node-aware SPS. Furthermore, when running real-world graph analytics applications and I/O-intensive pipelines, KubePACS delivers up to 23.8\% higher performance per dollar than Karpenter, with higher availability.

Our key contributions are summarized as follows.

\begin{itemize}[nosep, leftmargin=*]
\item The first attempt to use benchmark score, spot price, and multi-node SPS dataset together to solve a multi-objective optimization problem to build a large-scale compute cluster.
\item Proposing a workload-aware performance score adjustment heuristic for instances with specialized network and disk features.
\item Applying a GSS to identify the optimal cost-performance tradeoff with minimal operational overhead.
\item The open-source contribution for the research community.
\end{itemize}

\section{Spot Instance Key Considerations}
Spot instances utilize excess cloud resources to offer substantial cost savings~\cite{spotlake-iiswc}. However, availability fluctuations might result in node interruptions necessitating an intelligent provisioning strategy.

\subsection{Spot Instance Cost and Performance}
While on-demand pricing correlates with hardware specifications (e.g., CPU cores, memory, I/O), spot instance pricing is dynamically determined by supply and demand, often decoupling price from hardware performance. Thus, selection strategies based solely on minimal cost may provision inferior hardware, degrading execution times and increasing total costs.

Various benchmarking tools, such as Geekbench~\cite{geekbench}, SPEC~\cite{spec-benchmark}, and CoreMark~\cite{coremark}, evaluate computational performance. Notably, cloud providers adopt CoreMark scores~\cite{gcp-coremark, azure-coremark} as a holistic measure of capability beyond raw hardware specs. However, CPU metrics alone are insufficient, as non-CPU resources such as network and disk I/O bandwidths significantly impact workload performance. Unlike CPU and memory, I/O performance exhibits substantial variability depending on configurations and usage patterns, often not scaling proportionally with allocated bandwidth. Therefore, a comprehensive evaluation incorporating multiple dimensions of instance capabilities is essential for spot instance recommendation.

Figure~\ref{fig:coremark-and-price-variation} presents CoreMark scores (gray stars) on the primary vertical axis with on-demand (red diamonds) and spot prices (box-plots~\cite{box-whisker-plot}) on the secondary vertical axis across various AWS instance configurations. As shown in Figure~\ref{fig:instance-generation-coremark-price-variation}, newer generations (m6i to m8i) consistently yield higher computational performance, yet spot prices show a slight upward trend compared to stable on-demand costs. Figures~\ref{fig:instance-series-coremark-price-variation} and \ref{fig:instance-options-coremark-price-variation} highlight that on-demand prices vary significantly based on resource optimizations (e.g., memory, storage, or networking) even when CoreMark scores remain stable, indicating that higher costs from specialized hardware do not necessarily correlate with improved compute performance. Furthermore, Figure~\ref{fig:vendor-comparison-coremark-price} reveals that while on-demand price-to-performance ratios are consistent across CPU vendors, spot prices demonstrate distinct volatility patterns. These observations confirm that relying on spot price or CPU-centric metrics alone is insufficient for optimizing diverse workloads, as price is often decoupled from actual performance in the spot market.

\subsection{Spot Instance Availability}
Reliability is a crucial requirement in spot-based cluster provisioning. While early research relied on spot price volatility as a proxy for interruption risk~\cite{spot-analysis-javadi, deconstructing-spot-instance, draft-spot-instance-guarantee-from-spot-price, alibaba-spot-instance, stat-analysis-spot-price, spot-analysis-javadi, spot-instance-analysis, spot-price-by-location, spot-instance-for-hpc}, recent pricing policy changes have stabilized prices, decoupling them from real-time availability~\cite{spot-price-change-2017, spot-price-policy-change-2017-irwin}. Consequently, cloud providers introduced real-time availability metrics such as the SPS, offered by AWS~\cite{spot-placement-score-start} and Azure~\cite{azure-spot-placement-score}, ranging from 1 (Low) to 3 (High Availability).

However, existing tools often infer cluster-level availability based solely on single-node SPS~\cite{spotverse, spotlake-iiswc, single-node-sps-spot-interruption-visible}, leaving a critical gap in how these metrics are utilized. To illustrate the risks, we conducted a 24-hour experiment across four AWS regions, requesting 50 spot instances for two groups: one with a high SPS for 50 nodes, and another with a high SPS for a single node only.

\begin{figure}[t]
  \centering
  \includegraphics[width=0.48\textwidth]{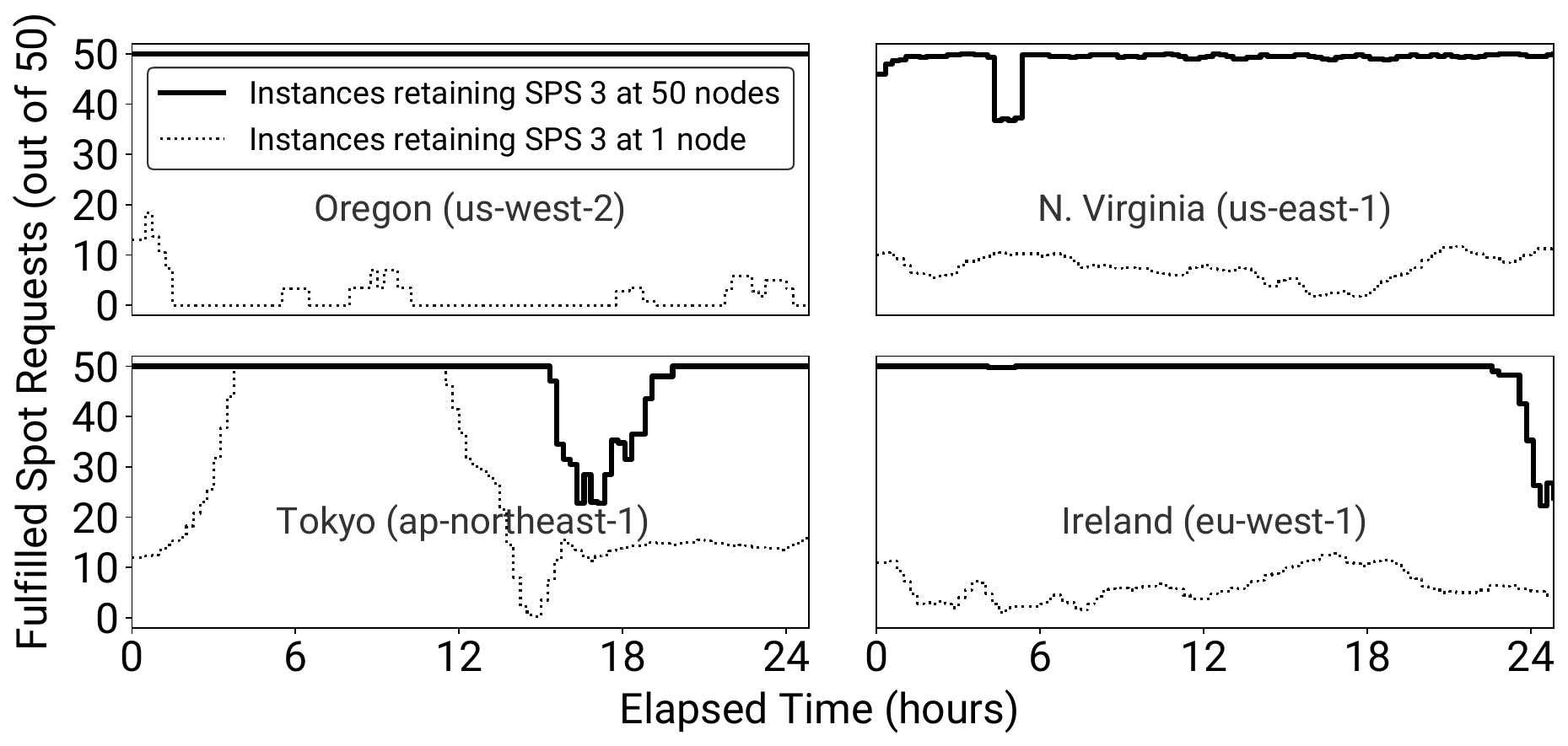}
  \caption{Different multiple spot instance availability for distinct single node and multi-node SPS values}
  \Description{Different multiple spot instance availability for distinct single node and multi-node SPS values}
  \label{fig:multiple-nodes-sps-real-availability}
\end{figure}

Figure~\ref{fig:multiple-nodes-sps-real-availability} empirically demonstrates the necessity of multi-node metrics. Instance types with a high multi-node SPS (solid lines) consistently maintained near-complete fulfillment (approx. 50 nodes), whereas those selected based solely on single-node SPS (dotted lines) exhibited severe provisioning failures, frequently fulfilling fewer than 10 instances despite high single-node SPS. This confirms that multi-node SPS is indispensable for large-scale spot node provisioning, a factor largely overlooked by prior approaches~\cite{spotverse, alibaba-spot-instance, stat-analysis-spot-price, draft-spot-instance-guarantee-from-spot-price} relying on spot prices or single-node metrics.

\subsection{Kubernetes Cluster With Spot Instances}
Kubernetes~\cite{kubernetes} has established itself as the standard container orchestration platform. Its dynamic pod allocation via the Horizontal Pod Autoscaler (HPA)~\cite{kubernetes-hpa}, node-level scaling via the Cluster Autoscaler (CA), and built-in fault-tolerance make it well-suited for batch workloads on cost-effective yet volatile spot instances.

Several systems and research projects integrate spot instances into clusters. Karpenter~\cite{karpenter} dynamically manages worker nodes by leveraging AWS SpotFleet's allocation strategies~\cite{spotfleet} for cost and availability optimization. Academic research, including SpotKube\allowbreak~\cite{spotkube}, SpotVerse~\cite{spotverse}, and others~\cite{stratus, eva, proteus, hotspot, tributary-atc18}, has explored techniques such as price prediction, completion time estimation, hybrid instance selection, and interruption-aware migration.

However, state-of-the-art approaches share a critical limitation: they neglect hardware performance heterogeneity. Existing schedulers typically provision nodes based solely on static resource requirements (e.g., vCPU count and memory size), implicitly assuming uniform performance across instance types. As shown in Figure~\ref{fig:coremark-and-price-variation}, this assumption leads to suboptimal selection, as instances with identical specifications can exhibit vast performance disparities. Furthermore, reliance on single-node SPS~\cite{spotverse} does not guarantee robustness for distributed deployments, as evidenced by Figure~\ref{fig:multiple-nodes-sps-real-availability}. Consequently, a provisioning mechanism that jointly optimizes cost, large-scale availability, and hardware performance is required.

\begin{figure}[t]
  \centering
  \includegraphics[width=0.47\textwidth]{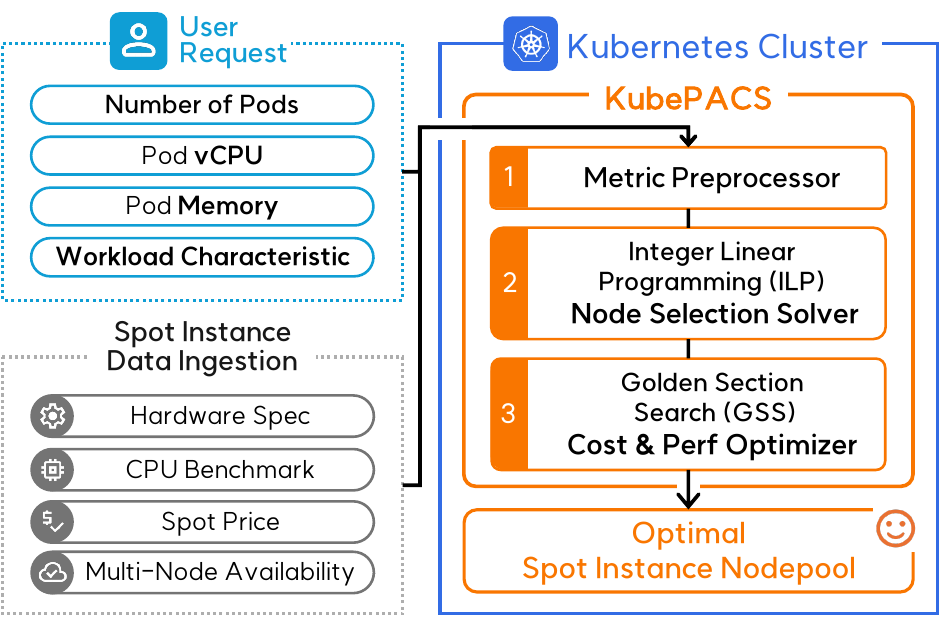}
  \caption{Overall system architecture of KubePACS}
  \Description{Overall system architecture of KubePACS}
  \label{fig:kubepacs-architecture}
\end{figure}

\section{KubePACS: System Architecture}
KubePACS is a framework designed to provision Kubernetes worker node pools that are \textbf{P}erformant, highly \textbf{A}vailable, and \textbf{C}ost-efficient by leveraging cloud \textbf{S}pot instances, resolving the multi-objective optimization problem of selecting optimal instance types under dynamic spot conditions. Figure~\ref{fig:kubepacs-architecture} depicts the overall architecture composed of data ingestion, metric processing, and optimization problem solving. The workflow initiates when a user submits workload requirements, including the total number of pods, per-pod resource specifications (vCPU and memory), and workload characteristics (e.g., I/O sensitivity). Simultaneously, the \emph{Spot Instance Data Ingestion} module aggregates real-time cloud datasets comprising hardware specifications, CPU benchmark scores (e.g., CoreMark~\cite{coremark}), spot prices, and multi-node availability metrics.

KubePACS processes inputs through a three-stage pipeline:
\begin{enumerate}[nosep, leftmargin=*]
  \item \textbf{Metric Preprocessor:} Synthesizes spot price, hardware benchmark scores, and multi-node SPS to compute allocatable pod counts and scale benchmark scores per workload requirements.
  \item \textbf{Integer Linear Programming (ILP) Node Selection Solver:} Formulates a multi-objective optimization problem to select instance configurations that satisfy resource constraints while optimizing the cost-performance balance (Section~\ref{sec:solver-design}).
  \item \textbf{Golden Section Search (GSS) Optimizer:} Iteratively tunes hyperparameters using the GSS algorithm~\cite{golden-section-method} to identify the optimal trade-off maximizing overall efficiency (Section~\ref{sec:solver-hyperparameter-search}).
\end{enumerate}

\begin{table}[!t]
  \centering
  \caption{Symbols used in the instance selection formulation}
  \begin{tabularx}{\linewidth}{lX}
    \toprule
    \textbf{Symbol} & \textbf{Description} \\
    \midrule
    $Req_{cpu}$     & The requested number of CPU cores per pod \\
    $Req_{mem}$     & The requested memory size per pod \\
    $Req_{pod}$     & The total number of requested pods \\
    $N$                 &The total number of candidate instance types \\
    $I_i$              & A candidate instance type with index $i$ \\
    $CPU_i$    & The number of CPU cores provided by $I_i$ \\
    $Mem_i$    & The memory size provided by $I_i$ \\
    $SP_i$    & Spot Price of $I_i$ \\
    $OP_i$    & On-demand Price of $I_i$ \\
    $BS_i$    & Single core Benchmark Score of $I_i$ \\
    $Pod_i$              & Number of pods that can be placed on $I_i$ \\
    $T3_i$             & Maximum number of $I_i$ spot instances with SPS of 3 \\
    $x_i$              & Number of $I_i$ instances to be allocated\\
    $Perf_{i}$         & Total Benchmark score of $I_i$, given by $BS_i \cdot Pod_i$  \\
    $\alpha$           & A hyperparameter for the cost-performance trade-off \\
    \bottomrule
  \end{tabularx}
  \label{tab:notation-summary}
\end{table}

Upon identifying the optimal configuration, the system provisions the spot instances via the cloud provider's SDK, integrating them into the Kubernetes cluster as worker nodes (Section~\ref{sec:implementation}).

To formulate the resource provisioning requirements, the user's request specification is defined as $Req$, comprising memory ($Req_{mem}$), CPU cores ($Req_{cpu}$), and the target number of pods ($Req_{pod}$). Uniform-sized pods are assumed to facilitate ease of scaling for general workloads~\cite{pywren, hadoop, cloud-usage-pattern} in the cloud. Even if multiple workloads with different pod specifications are submitted concurrently, KubePACS optimizes the node pool for each independently, enabling diverse configurations within a single cluster.

Given user preferences (e.g., instance category, region), a set of $N$ candidate instance types is identified. Each candidate instance type, $I_i$ ($0 \leq i < N$), represents a unique instance type within a specific AZ to account for distinct spot prices, denoted as $SP_i$. The allocatable CPU cores and memory of $I_i$ are defined as $CPU_i$ and $Mem_i$, respectively. To quantify computational performance, the CoreMark benchmark score~\cite{coremark} is adopted and denoted as $BS_i$.

Let $x_i$ be the number of provisioned instances of type $I_i$. The number of pods allocatable to $I_i$, denoted as $Pod_i$, is derived in Equation~\ref{eq:pod-i-definition} to satisfy constraints. Notations are summarized in Table~\ref{tab:notation-summary}.

\begin{equation}\label{eq:pod-i-definition}
  Pod_i = \min \left(
    \left\lfloor \frac{CPU_i}{Req_{cpu}} \right\rfloor,
    \left\lfloor \frac{Mem_{i}}{Req_{mem}} \right\rfloor
  \right)
\end{equation}

To optimize cluster composition, two efficiency metrics are introduced. First, \textit{performance-cost efficiency} ($E_{PerfCost}$) represents the cumulative performance-per-dollar of selected instances. While maximizing $E_{PerfCost}$ promotes the selection of high-performance and cost-effective instances, doing so without constraint may cause over-provisioning by selecting more instances than necessary, increasing the overall cost. To address this issue, the \textit{excess pod allocation efficiency} ($E_{OverPods}$) quantifies the ratio of requested pods to the total allocatable capacity, penalizing over-provisioning. Metrics are formalized in Equation~\ref{eq:efficiency-cost-perf-over}. Given the nature of the cloud, allocatable capacity is assumed to satisfy the request, i.e., $E_{OverPods} \le 1.0$.

\begin{equation}\label{eq:efficiency-cost-perf-over}
  E_{PerfCost} = \sum_{i}\frac{BS_i \cdot x_i}{SP_i}, \quad
  E_{OverPods} = \frac{Req_{pod}}{\sum\limits_{i}Pod_i \cdot x_i}
\end{equation}

KubePACS aims to recommend a set of spot instances including instance types and their counts that maximizes the \textit{total efficiency}, $E_{Total}$, defined as in Equation~\ref{eq:efficiency-overall}.

\begin{equation}\label{eq:efficiency-overall}
  E_{Total} = E_{PerfCost} \times E_{OverPods}
\end{equation}

In this formulation, $E_{PerfCost}$ encourages the use of spot instances with high benchmark scores and significant cost savings, while $E_{OverPods}$ penalizes solutions that result in excessive resource allocation beyond the workload's requirements. This combination enables the system to balance performance with cost efficiency.

\subsection{Optimal Node Selection Solver Design}\label{sec:solver-design}
Assigning pods with specific resource requirements to spot instances can be formulated as a bin-packing problem~\cite{bin-packing-algorithm}. However, the dynamic price and heterogeneous performance of spot instances necessitate a joint optimization of cost-efficiency and computational power, extending the problem beyond the traditional bin-packing problem. Consequently, this work formulates the allocation strategy as an Integer Linear Programming (ILP) problem to handle multi-dimensional constraints optimally.

The total performance contribution of an instance $I_i$ is defined as $Perf_i = BS_i \times Pod_i$, representing the aggregate benchmark score for all hosted pods. To resolve scale discrepancies between large benchmark scores and small hourly costs when solving an ILP problem, Min-Normalization is applied to both metrics, with the minimum of each metric defined in Equation~\ref{eq:min-scores}, chosen for its demonstrated effectiveness in multi-objective scaling~\cite{normalize}.

\begin{equation}\label{eq:min-scores}
  Perf_{\min} = \min_{0\le i < N} \left( BS_i \cdot Pod_i \right), \quad
  SP_{\min} = \min_{0\le i < N} \left( SP_i \right)
\end{equation}

To maximize $E_{Total}$, the ILP solver must account for the pod over-allocation factor $E_{OverPods}$ during objective function evaluation. However, $E_{OverPods}$ can only be computed after solving the allocation problem, introducing a cyclic dependency that renders the objective function unsolvable using standard ILP solvers. A naive approach of enforcing strict equality between allocated and requested pods restricts the solution space, potentially precluding configurations where slight over-provisioning utilizes cheaper, larger instances to improve overall efficiency.

To mitigate this, the objective function is formulated to determine the optimal instance count $x_i$ by minimizing a weighted sum of normalized performance and cost, as presented in Equation~\ref{eq:objective-function}.

\begin{equation}\label{eq:objective-function}
  \begin{gathered}
    \displaystyle \text{minimize} \sum_{i} \left(
      - \alpha \cdot \frac{Perf_i}{Perf_{min}}
      + (1 - \alpha) \cdot \frac{SP_i}{SP_{min}}
    \right) \cdot x_i \\
    \text{subject to:} \quad x_i \leq T3_i, \quad x_i \in \mathbb{Z}_{\ge 0}
  \end{gathered}
\end{equation}

The formulation introduces a tunable hyperparameter $\alpha \in [0.0,\\ 1.0]$ to balance the trade-off between maximizing performance and minimizing cost. A higher $\alpha$ prioritizes performance, which potentially leads to over-provisioning, while a lower $\alpha$ favors strict cost reduction. To guarantee robust availability of selected spot instances, the constraint $x_i \leq T3_i$ leverages the multi-node SPS dataset~\cite{spotlake-iiswc, multi-spotlake-www, multinode-spot-dataset}. Specifically, $T3_i$ is defined as the maximum number of simultaneous instances for type $I_i$ that maintain an SPS of 3 (highest availability). Given the non-increasing nature of SPS with respect to request size~\cite{multinode-spot-dataset}, limiting allocations to $T3_i$ ensures that the provisioned cluster operates with high availability, thereby minimizing interruption risks.

\subsection{Cost-Performance Hyperparameter Tuning}\label{sec:solver-hyperparameter-search}
The optimal instance configuration minimizing Equation~\ref{eq:objective-function} is highly sensitive to the weight parameter $\alpha$. Low $\alpha$ values prioritize cost reduction, restricting node counts to the minimum required. Conversely, high $\alpha$ values emphasize performance, potentially justifying over-provisioning if the performance gain outweighs the marginal cost increase. Consequently, identifying the specific $\alpha$ that maximizes the total efficiency metric, $E_{Total}$, is essential.

To identify this optimal $\alpha$, we employ the Golden Section Search (GSS) algorithm~\cite{golden-section-method}, a classical method for efficiently locating the maximum of a unimodal function. GSS iteratively narrows the search interval using the golden ratio, $\phi = \frac{\sqrt{5} - 1}{2} \approx 0.618$, by selecting two interior points that divide the interval proportionally. The point with the lower function value is discarded at each step, contracting the interval by a factor of approximately $\phi$. GSS is chosen for its superior convergence rate compared to alternatives like ternary search. By reusing intermediate function evaluations, GSS requires only one objective function evaluation per iteration after initialization~\cite{golden-section-equation}. This characteristic minimizes the computational overhead of the iterative ILP solving process.

The search operates within the range $\alpha \in [0.0, 1.0]$ and terminates when the interval width falls below a tolerance $\varepsilon = 10^{-n}$. The number of iterations, $k$, required to guarantee this precision is derived from the contraction factor $\phi \approx 0.618$~\cite{golden-section-equation}. The general bound of $k$ is defined as follows.

\begin{equation}
  k-1 \geq \left\lceil \frac{\log(\varepsilon / (b - a))}{\log(\phi)} \right\rceil
\end{equation}

Given $(b - a) = 1.0$ and $\varepsilon = 0.1^n$, we obtain:

\begin{equation}\label{eq:alpha-tolerance-range}
  k - 1\geq \left\lceil \frac{\log(10^{-n})}{\log(0.618)} \right\rceil
  = \left\lceil \frac{-n \cdot \log(10)}{\log(0.618)} \right\rceil
  \approx \left\lceil 4.784 \cdot n \right\rceil
\end{equation}

Approximately $5n + 1$ iterations are required to achieve the tolerance $\varepsilon$, with the search range tolerance shrinking exponentially as the number of iterations increases linearly with $n$. Empirical analysis (Figure~\ref{fig:impact-of-alpha-spacing}) suggests that $n=2$, a tolerance of 0.01, achieves an optimal balance, locating the target $\alpha$ with negligible overhead.

\subsection{Workload-Aware Performance Scaling}\label{sec:io-intensive-task-scaling}

While CoreMark robustly measures compute-bound performance, it fails to capture the performance benefits of specialized network or storage hardware. As shown in Figure~\ref{fig:coremark-and-price-variation}, cloud providers impose price increments for such features. Consequently, the optimization solver (Equation~\ref{eq:objective-function}) would penalize these instances due to elevated costs ($SP_i$) and identical CPU benchmarks ($BS_i$), despite their suitability for I/O-bound applications.

To address this discrepancy, a scaling mechanism is applied when a user specifies a workload preference (e.g., \emph{network-} and/or \emph{disk-intensive}). For instance types matching the requested capability, the benchmark score $BS_i$ is adjusted using the ratio of their on-demand price to the base instance price.

\begin{equation}
  BS_i^{scaled} = BS_i \times \frac{OP_i}{OP_{base}}
\end{equation}

Here, $OP_i$ denotes the on-demand price of specialized instance $I_i$, and $OP_{base}$ the price of the corresponding general-purpose instance in the same family. This approach leverages the cloud provider's pricing model, implicitly quantifying the value of hardware enhancements where direct benchmarking is impractical.

For example, in a \emph{network-intensive} scenario, the score of a \emph{c6in} (network-optimized) instance is scaled up by the ratio of its price, $\$0.23$, to the base \emph{c6i} of $\$0.17$, effectively increasing its competitiveness by $\frac{0.23}{0.17}$ in the solver. Conversely, non-matching types like \emph{c6id} (disk-optimized) remain unscaled. These adjusted scores are subsequently integrated into the objective function to ensure appropriate resource selection.

When no workload preference is specified, the workload-aware scaling is not applied and all candidate instances are evaluated with uniform benchmark weighting. Even if an incorrect preference is provided, the system provisions a fully functional cluster, as only the hardware specialization scoring is affected without compromising availability or correctness.

\begin{algorithm}[t!]
  \caption{KubePACS Node Selection}
  \label{algo:procedure}
  \begin{algorithmic}[1]
    \State \textbf{Input:} Pod spec$(Req_{pod}, Req_{\text{cpu}}, Req_{\text{mem}})$, workload $W$
    \State \textbf{Output:} Node pool configuration $\{(I_i, x_i)\}$ satisfying total pod requirement
    \State Initialize $\mathcal{I} \gets \emptyset$
    \For{each instance type $I_i$}
    \State $\mathcal{I} \gets \mathcal{I} \cup$ DatasetPreProcessing($I_i$, {$W$}, {$Req_{\text{cpu}}$}, {$Req_{\text{mem}}$})
    \EndFor
    \State Initialize search interval: $\alpha_{\text{left}} \gets 0.0$, $\alpha_{\text{right}} \gets 1.0$
    \State $\phi \gets 0.618$ \Comment{Golden ratio}
    \State $(\alpha_1, \alpha_2) \gets (\alpha_{\text{right}} - \phi,\ \alpha_{\text{left}} + \phi)$
    \State $\mathcal{S}_1 \gets \texttt{ILP}(\alpha_1, \mathcal{I}, Req_{pod}),\ \mathcal{S}_2 \gets \texttt{ILP}(\alpha_2, \mathcal{I}, Req_{pod})$
    \State $\mathcal{S}^* \gets \text{argmax between }(\mathcal{S}_1,\ \mathcal{S}_2) \text{ by } E_{Total}$
    \While{$\alpha_{\text{right}} - \alpha_{\text{left}} > \varepsilon$}
    \If{$E_{Total}(\mathcal{S}_1) \ge E_{Total}(\mathcal{S}_2)$}
    \State $\alpha_{\text{right}} \gets \alpha_2$
    \State $\alpha_2 \gets \alpha_1$, $\mathcal{S}_2 \gets \mathcal{S}_1$
    \State $\alpha_1 \gets \alpha_{\text{right}} - \phi \cdot (\alpha_{\text{right}} - \alpha_{\text{left}})$
    \State $\mathcal{S}_1 \gets \texttt{ILP}(\alpha_1, \mathcal{I}, Req_{pod})$
    \State $\mathcal{S}^* \gets \text{argmax between } (\mathcal{S}^*,\ \mathcal{S}_1) \text{ by } E_{Total}$

    \Else
    \State $\alpha_{\text{left}} \gets \alpha_1$
    \State $\alpha_1 \gets \alpha_2$, $\mathcal{S}_1 \gets \mathcal{S}_2$
    \State $\alpha_2 \gets \alpha_{\text{left}} + \phi \cdot (\alpha_{\text{right}} - \alpha_{\text{left}})$
    \State $\mathcal{S}_2 \gets \texttt{ILP}(\alpha_2, \mathcal{I}, Req_{pod})$
    \State $\mathcal{S}^* \gets \text{argmax between } (\mathcal{S}^*,\ \mathcal{S}_2) \text{ by } E_{Total}$
    \EndIf
    \EndWhile
    \State \Return Solution $\mathcal{S}^*$ with highest $E_{Total}$
  \end{algorithmic}
\end{algorithm}

Algorithm~\ref{algo:procedure} outlines the procedure for generating a cost-efficient and reliable worker node configuration, comprising two stages.

In the first stage (Lines 3--6), the system aggregates and normalizes instance-level metadata. The function \emph{DatasetPreProcessing} takes a candidate instance $I_i$, the workload characteristics $W$, and per-pod resource requirements ($Req_{\text{cpu}}, Req_{\text{mem}}$) as inputs. It computes the maximum allocatable pod count ($Pod_i$) for the instance and scales the benchmark score ($BS_i$) according to the user-defined workload preferences (e.g., disk and/or network heavy). The output is an enriched dataset $\mathcal{I}$ containing all necessary attributes for optimization.

In the second stage, starting from Line 7, the system executes a hyperparameter optimization over the cost-performance weight $\alpha \in [0.0, 1.0]$ using the GSS algorithm to maximize $E_{Total}$. For each candidate $\alpha$, the solver function \emph{ILP} is invoked to generate a candidate node pool $\mathcal{S}$. This function formulates the selection task as an ILP problem (Equation~\ref{eq:objective-function}), determining the optimal set of instances that satisfies the total pod demand ($Req_{pod}$) and availability constraints ($T3_i$). The solver aims to jointly minimize cost and over-allocation while maximizing hardware performance under the specified $\alpha$. The GSS algorithm iteratively refines $\alpha$, ultimately returning the configuration $\mathcal{S}^*$ that yields the highest overall efficiency among the evaluated candidates.

\section{KubePACS Implementation}\label{sec:implementation}

\begin{figure}[t]
  \centering
  \includegraphics[width=0.48\textwidth]{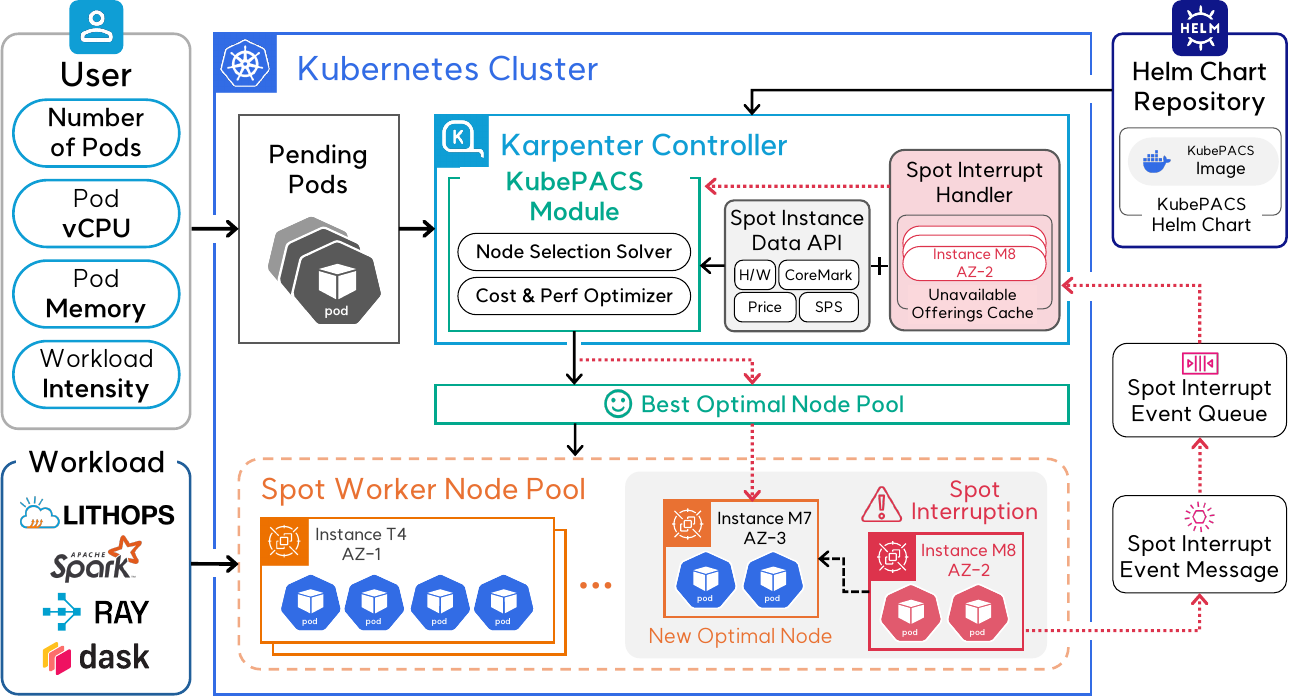}
  \caption{Implementation of KubePACS to provision Kubernetes worker nodes}
  \Description{Implementation of KubePACS to provision Kubernetes worker nodes}
  \label{fig:kubepacs-k8s}
\end{figure}

KubePACS is implemented as a Python-based prototype module integrated directly into the \textit{Karpenter Controller} to facilitate Kubernetes-native node auto-scaling. The overall workflow is shown in Figure~\ref{fig:kubepacs-k8s}. The system utilizes a forked codebase of Karpenter~\cite{karpenter}, intercepting the standard node expansion workflow triggered by \textit{Pending Pods} to inject the proposed KubePACS instance selection logic. To construct an optimal node pool that maximizes the proposed objective function, $E_{Total}$, the problem is modeled and solved using the Python \texttt{PuLP} library (v.3.0.2)~\cite{pulp}, a linear programming modeler.

Upon determining the optimal configuration, the \textit{Node Selection Solver} transmits the solution to the Karpenter controller.
Karpenter then provisions the \textit{Spot Worker Node Pool} using the recommended instance types. New nodes subsequently join the cluster, allowing the pending pods to be rapidly scheduled.

\subsection{Spot Interruption Handling Mechanism}
A critical component of the KubePACS implementation is its robust mechanism for handling spot instance volatility. As depicted in Figure~\ref{fig:kubepacs-k8s}, spot interruption notifications from the cloud provider are captured as \textit{Spot Interrupt Event Messages} and forwarded to a \textit{Spot Interrupt Event Queue}. The \textit{Spot Interrupt Handler} asynchronously processes these events and identifies the interrupted instance types. These interrupted offerings are immediately recorded in the \textit{Unavailable Offerings Cache}. During the subsequent re-optimization cycle, the \textit{Node Selection Solver} queries this cache to enforce constraints that exclude unstable offerings. Consequently, the system provisions a \textit{New Optimal Node} that maximizes $E_{Total}$ while avoiding the interrupted availability zone or instance type.  This reactive loop ensures rapid capacity recovery and maintains workload continuity.

\subsection{Deployment and Distribution}
To ensure ease of deployment and seamless integration into existing Kubernetes environments, KubePACS is packaged as a Docker container image and distributed via a standard \textit{Helm Chart Repository} as the \textit{KubePACS Helm Chart}. This packaging strategy allows administrators to deploy KubePACS alongside the Karpenter scheduler with minimal configuration overhead. To facilitate reproducibility and further research, the complete source code, Helm charts, and deployment artifacts are made publicly available.\footnote{https://kubepacs.ddps.cloud}

\section{Evaluation}
\label{sec:evaluation}
The implemented framework is empirically evaluated to address the following research questions.
\\
\begin{itemize}[nosep, leftmargin=*, itemsep=4pt]
{
\item\textbf{RQ-1 (Comparative Analysis):} How does the KubePACS recommendation algorithm compare to state-of-the-art approaches (e.g., SpotVerse~\cite{spotverse}, SpotKube~\cite{spotkube}) in terms of cost-efficiency, hardware performance, and availability?

\item\textbf{RQ-2 (Design Validation):} How do internal design choices of KubePACS, specifically the cost-performance hyperparameter ($\alpha$) and workload-aware scaling, impact the system's effectiveness?

\item\textbf{RQ-3 (Real-world Impact):} To what extent does KubePACS improve performance and reduce costs compared to a production-grade Kubernetes baseline, Karpenter~\cite{karpenter} with AWS SpotFleet?
}
\end{itemize}
\begin{figure*}[t]
  \centering
  \includegraphics[width=0.7\textwidth]{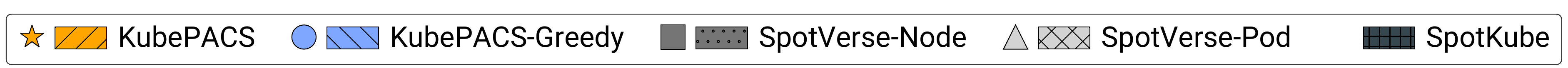}
  \subfloat[Overall efficiency ($E_{Total}$)]{
    \includegraphics[height=0.21\textwidth]{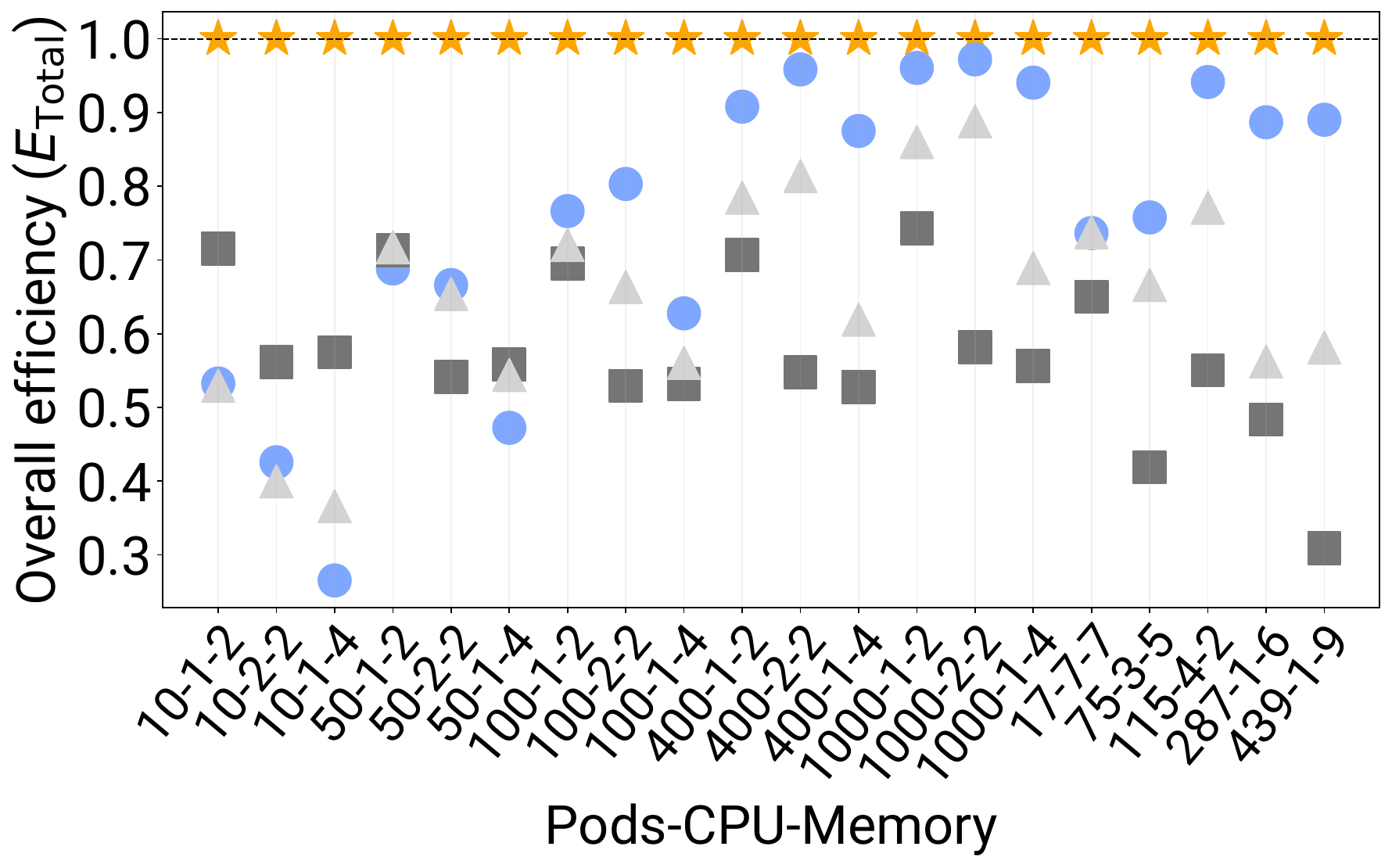}
    \label{fig:comparison-related-work-eff}
  }
  \hfill
  \subfloat[Spot instance availability metric]{
    \includegraphics[height=0.21\textwidth]{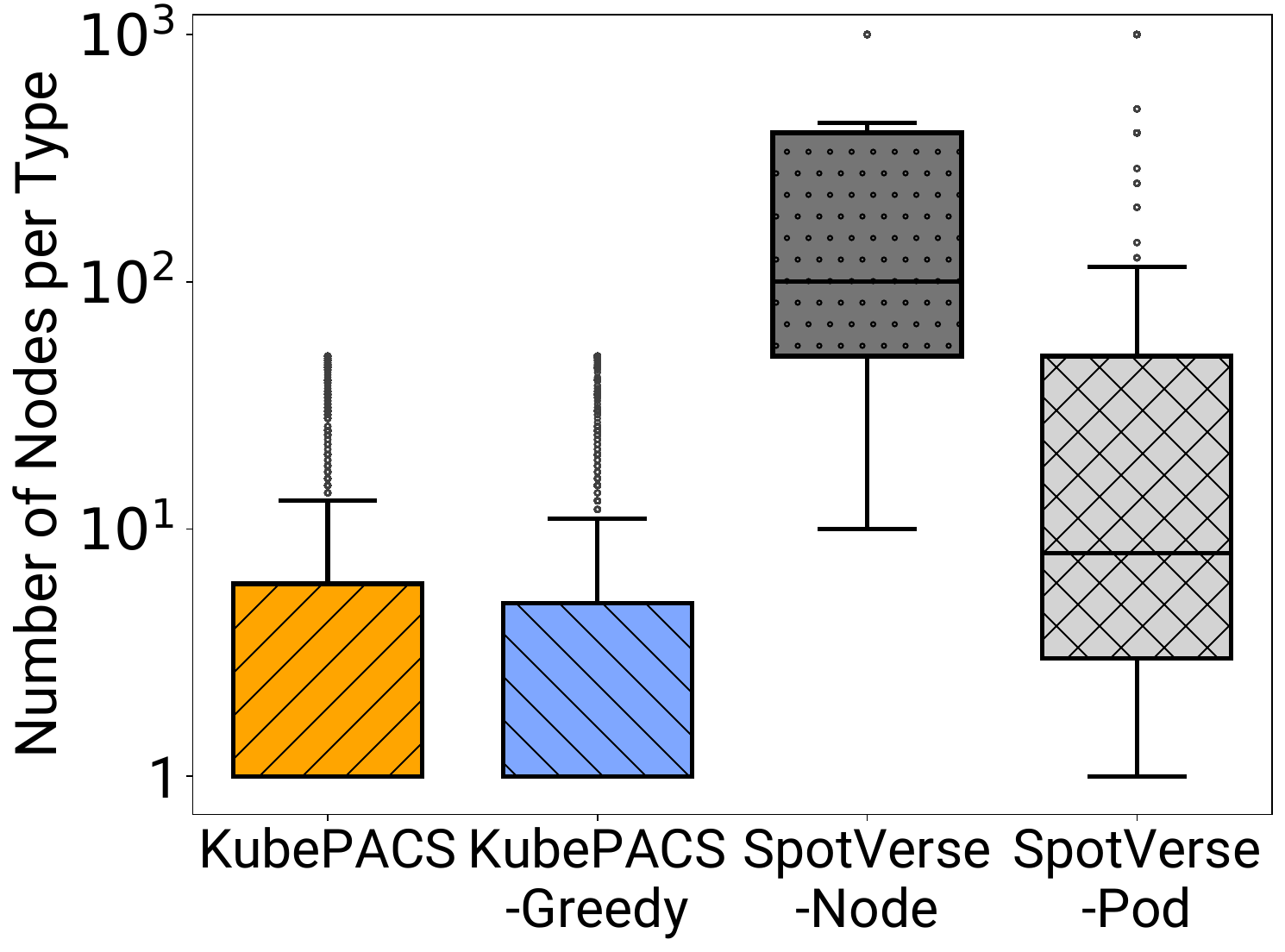}
    \label{fig:compare-max-node-t3}
  }
  \hfill
  \subfloat[Small-scale Microservice]{
    \includegraphics[height=0.21\textwidth]{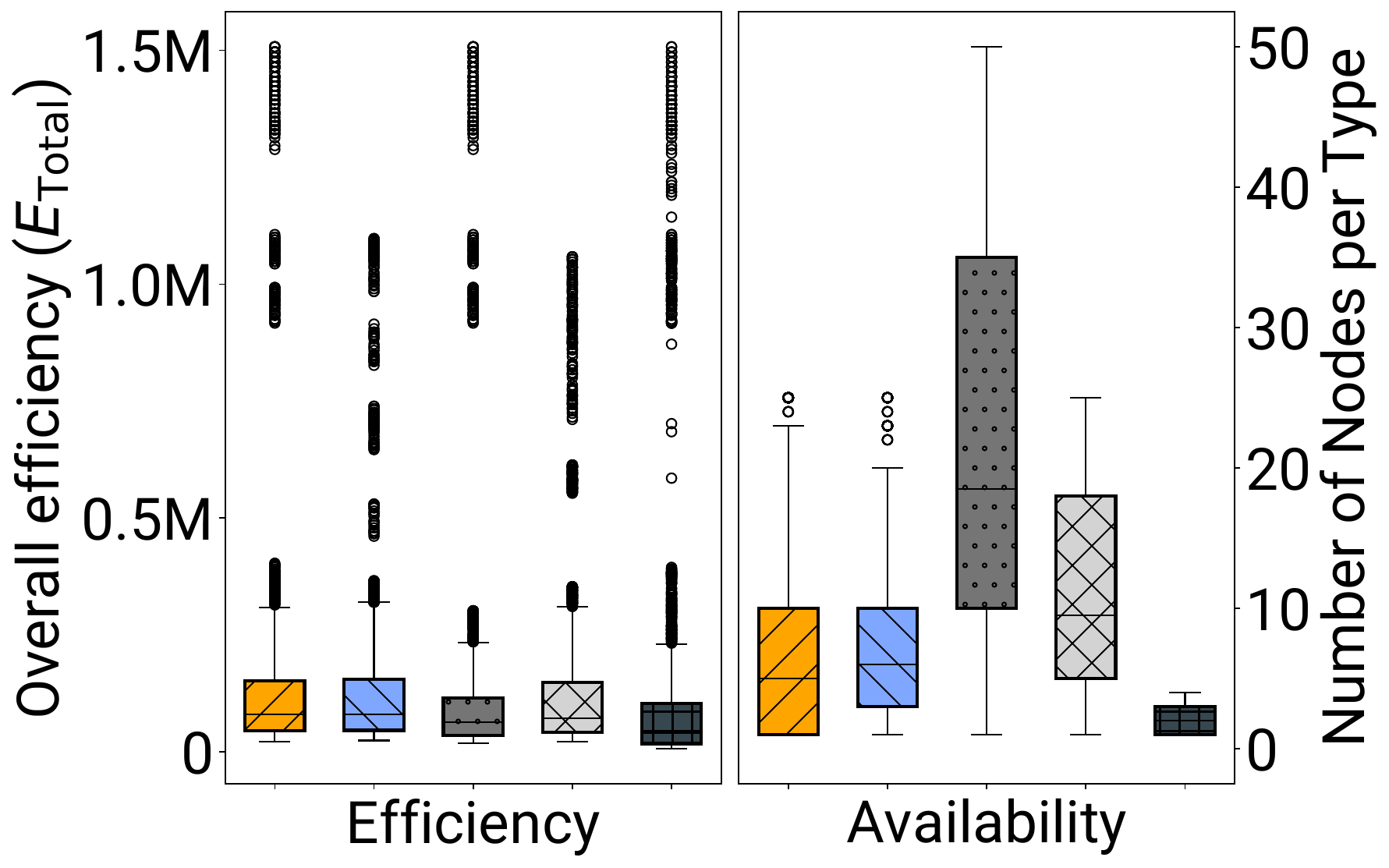}
    \label{fig:comparison-msa}
  }\hfill
  \caption{Comparing KubePACS with related works shows superb performance for cost and performance efficiency}
  \label{fig:comparison-related-work}
\end{figure*}

\subsection{Experiment Setup}

\paragraph{Kubernetes Cluster Configuration Scenarios}
To diversify the evaluation, we synthetically generate 15 Kubernetes cluster composition scenarios using the Cartesian product of requested pod counts \{10, 50, 100, 400, 1000\} and pod configurations \{(1 vCPU, 2 GiB), (2 vCPU, 2 GiB), (1 vCPU, 4 GiB)\}. Each scenario is represented as a tuple \texttt{(Number of pods, vCPU, Memory)}. Additionally, we include 5 irregular configurations—\{(17, 7, 7), (75, 3, 5), (115, 4, 2), (287, 1, 6), (439, 1, 9)\}—resulting in a total of 20 scenarios.

\paragraph{Testbed Environment and Metrics}
Datasets comprising spot and on-demand prices, benchmark scores, and SPS (both single- and multi-node) are acquired via SpotLake~\cite{spotlake-iiswc}. The collection period spans November 1--15, 2025, encompassing 731 instance types across all AZs in four AWS regions: N. Virginia, Oregon, Ireland, and Tokyo. For comparative analysis against a production-grade baseline, an Amazon EKS cluster is provisioned using Karpenter (v1.4)~\cite{karpenter}, with experiments conducted over a 12-hour window on May 22 in the corresponding regions. To simulate realistic application scenarios, Lithops~\cite{lithops-tcc} is employed to orchestrate compute-, network-, and disk-intensive tasks within the Kubernetes environment. The quantitative analysis focuses on the total hourly cluster cost and the efficiency metrics of $E_{PerfCost}$, $E_{OverPods}$, and $E_{Total}$.

\begin{figure*}[t!]
  \centering
  \subfloat[AWS]{
    \includegraphics[width=1.08\columnwidth]{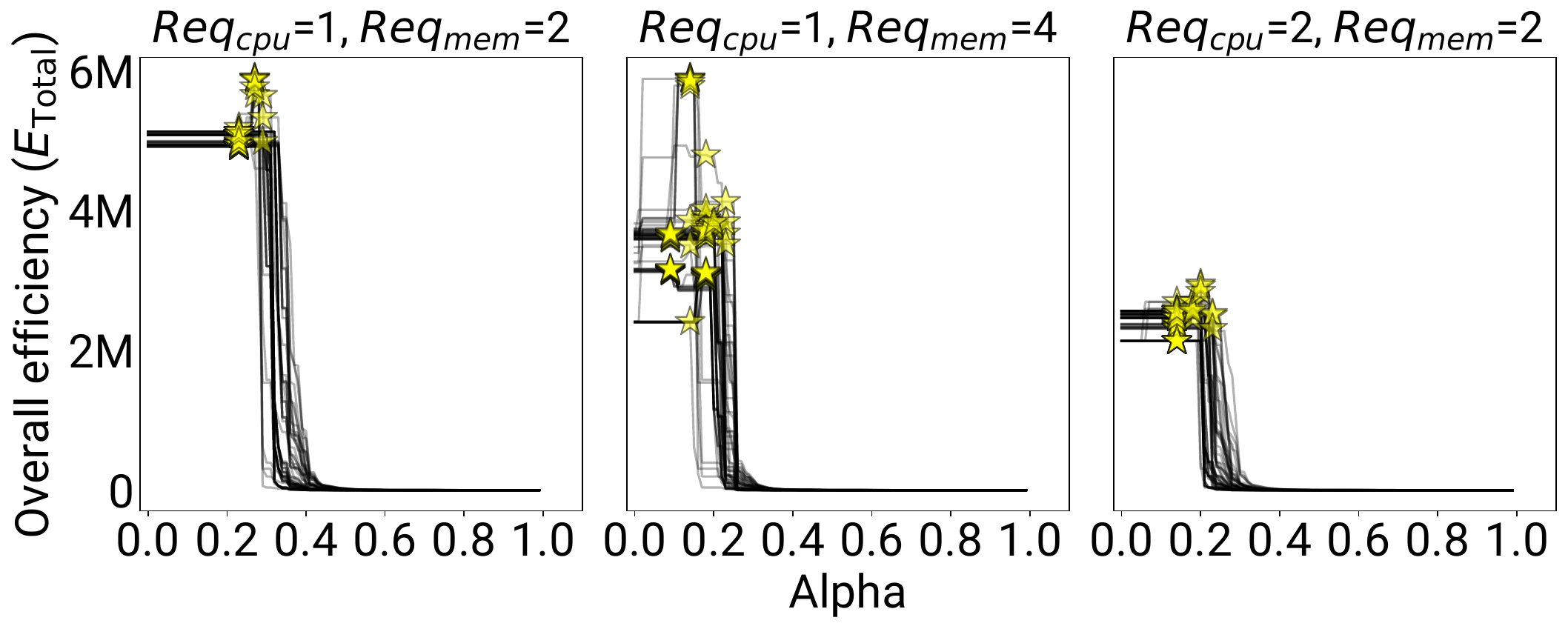}
    \label{fig:distribution-of-alphas-aws}
  }
  \subfloat[Azure]{
    \includegraphics[width=1.01\columnwidth]{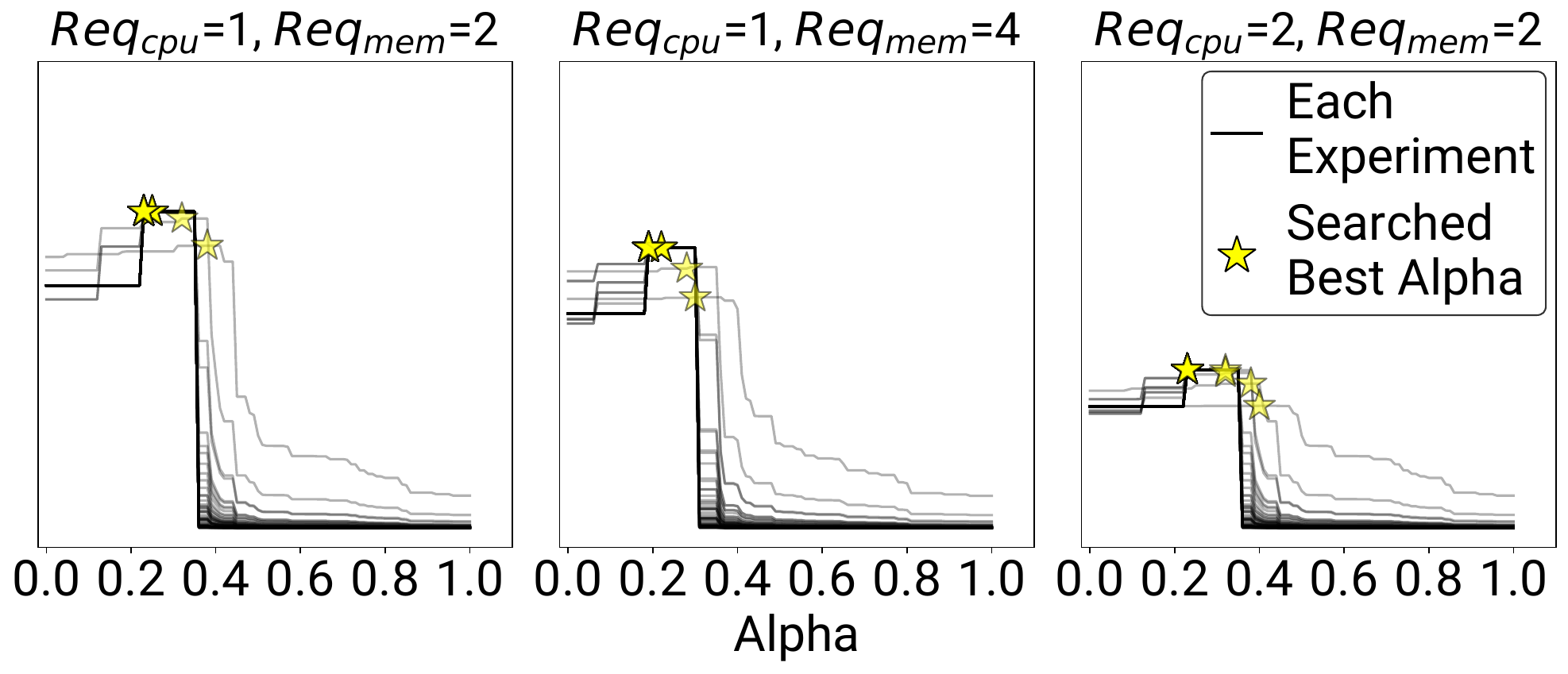}
    \label{fig:distribution-of-alphas-azure}
  }
  \hfill
  \caption{Overall efficiency ($E_{Total}$) changes with $\alpha$, the cost-performance trade-off parameter}
  \Description{Overall efficiency ($E_{Total}$) changes with $\alpha$, the cost-performance trade-off parameter}
  \label{fig:distribution-of-alphas-to-workloads}
\end{figure*}

\subsection{Comparing KubePACS with State-of-the-Art}\label{sec:comparing-to-related-work}

To address \textbf{RQ-1}, KubePACS is evaluated against the following baselines in terms of cost-efficiency, performance, and availability.

\emph{KubePACS-Greedy} serves as an ablation baseline, utilizing the identical dataset of KubePACS but employing a naive allocation strategy. Candidate instances are ranked by performance-cost efficiency ($E_{PerfCost}$), and pods are greedily allocated to top-ranked instances with the $T3$ constraint until the demand is met.

\emph{SpotVerse}~\cite{spotverse} selects candidates based on spot price, single-node SPS, and Interruption Frequency (IF). It filters out instance–region pairs whose combined SPS and IF score exceeds a threshold, prioritizing lower-priced options. Since SpotVerse originally operates at the instance level, two variants are adapted to align with Kubernetes pod semantics: \textit{SpotVerse-Node} (lowest price per node) and \textit{SpotVerse-Pod} (lowest price per pod).

\emph{SpotKube}~\cite{spotkube} employs an NSGA-II~\cite{genetic-algorithm} genetic algorithm to optimize deployments. It utilizes a Pareto-based fitness function to balance cost against availability, enhancing resilience by distributing pods across diverse instance types and AZs.

Figure~\ref{fig:comparison-related-work} presents a comparative analysis of KubePACS and related approaches. In Figure~\ref{fig:comparison-related-work-eff}, the y-axis represents the normalized overall efficiency ($E_{Total}$) across various pod requirement scenarios on the x-axis. The efficiency values are normalized relative to KubePACS, where a value of 1.0 serves as the reference; values below 1.0 indicate lower efficiency than KubePACS. Experimental results demonstrate that KubePACS outperforms all baselines, achieving average efficiency improvements of 48.11\%, 81.06\%, and 60.40\% compared to \emph{KubePACS-Greedy}, \emph{SpotVerse-Node}, and \emph{SpotVerse-Pod}, respectively. The reduced efficiency observed in \emph{KubePACS-Greedy} is primarily attributed to excessive pod over-allocation, which negatively impacts $E_{OverPods}$. In contrast, the SpotVerse variants prioritize price and availability, failing to consider instance performance. Notably, \emph{SpotVerse-Pod} exhibits substantial over-allocation, whereas \emph{SpotVerse-Node} allocates fewer pods per node.

Figure~\ref{fig:compare-max-node-t3} evaluates the availability implications of spot instance recommendations generated by each approach. To estimate the stability of spot instance requests, the number of allocated nodes per instance type is analyzed. This metric is motivated by the observation that excessive reliance on a single instance type significantly increases the risk of simultaneous interruptions, thereby reducing overall reliability~\cite{spotkube}. The vertical axis represents the number of nodes allocated to each instance type, visualized using a box-and-whisker plot on a logarithmic scale. The experimental dataset comprises 4,800 samples collected over a 15-day period across four AWS regions, covering 20 distinct scenarios at six-hour intervals.

As illustrated in the figure, KubePACS effectively constrains the number of instances per type by utilizing the $T3$ metric as a strict upper bound. In contrast, SpotVerse does not impose such a constraint and frequently concentrates allocations onto a single instance type. This tendency is particularly pronounced in the \emph{SpotVerse-Node} variant, which often recommends hundreds of identical instances. This lack of diversity exacerbates the risk of correlated failures, negatively impacting the aggregate availability of the cluster.

\paragraph{Comparison with SpotKube in Small-scale Scenarios.}
SpotKube was omitted from the extensive large-scale experiments as its original evaluation framework~\cite{spotkube} is specifically designed for small-scale microservice environments. To ensure a fair baseline comparison, the experimental setup described in the SpotKube publication~\cite{spotkube} was replicated. The workload involved pod counts ranging from 1 to 50, with each pod configured to require 1 vCPU and 1 GiB of memory. The candidate node pool was restricted to the specific instance types of \allowbreak\emph{t3.medium}, \allowbreak\emph{c6a.large}, \allowbreak\emph{t4g.large}, and \allowbreak\emph{c6g.xlarge}.

Figure~\ref{fig:comparison-msa} presents the comparative results, with the left chart depicting overall efficiency ($E_{Total}$) and the right showing allocated instances per type. KubePACS achieves the highest efficiency, approximately 107\% higher than SpotKube, primarily because SpotKube's rigid reliability mechanism enforces a fixed count of four instances per type, often forcing the selection of less efficient nodes to satisfy instance type diversity. In contrast, KubePACS employs a dynamic instance cap based on the $T3$ metric, effectively balancing cost-performance efficiency with robust spot availability guarantees.

Beyond KubePACS's superior efficiency, \emph{KubePACS-Greedy} exhibits comparable performance, attributable to the limited cardinality of the candidate instance pool, which leads both algorithms to converge on similar recommendation sets. This highlights KubePACS's strength in large-scale public cloud environments, where its rigorous optimization becomes essential over greedy heuristics that suffice only for small and constrained search spaces.

\subsection{Analysis of Internal KubePACS Mechanisms}
This section analyzes the internal operational characteristics of KubePACS to address \textbf{RQ-2}.

\paragraph{Impact of Cost-Performance Weight Parameter ($\alpha$).}
Figure~\ref{fig:distribution-of-alphas-to-workloads} illustrates the variation of overall efficiency ($E_{Total}$) with respect to $\alpha$ during the GSS algorithm's exploration. The data were collected from 12 independent runs at 6-hour intervals between November 3--5, 2025, in AWS N. Virginia{, and between February 16--18, 2026, in Azure US East}. Black lines represent efficiency trajectories of explored node pools across varying $\alpha$ values, while yellow stars denote the $\alpha$ yielding maximum $E_{Total}$ in each run.

As $\alpha$ increases from 0.0, which prioritizes spot cost exclusively, $E_{Total}$ initially rises due to the selection of instances with superior hardware performance at marginal cost increase. However, exceeding the optimal $\alpha$ threshold causes a sharp decline in $E_{Total}$, resembling a step-down function, primarily driven by excessive pod over-allocation penalizing $E_{OverPods}$. Compared to the $\alpha=0$ baseline mirroring cost-centric approaches, optimizing $\alpha$ improves $E_{Total}$ by an average of 6\% and up to 81\%, demonstrating the advantage of incorporating hardware metrics into node recommendation.

The same experiment on Azure, shown in Figure~\ref{fig:distribution-of-alphas-azure}, confirms cross-provider generalizability, as the characteristic concave pattern of $E_{Total}$ with respect to $\alpha$ is consistently observed. However, Azure's absolute $E_{Total}$ values are approximately 15\% lower than AWS, attributable to limited SPS data coverage and a lower number of available instance types: only 17.9\% of candidate instance types maintained consistently valid SPS during the experimental period. Accordingly, the remaining experiments are conducted on AWS, where comprehensive spot instance data is available.

\begin{table}[t]
  \centering
  \footnotesize
  \setlength{\tabcolsep}{8pt}
  \caption{Normalized $E_{Total}$ comparison across configurations}
  \begin{tabularx}{\columnwidth}{>{\centering\arraybackslash}Xccccc}
    \toprule
    & \textbf{Greedy} & $\boldsymbol{\alpha=0}$ & $\boldsymbol{\alpha=0.5}$ & $\boldsymbol{\alpha=1.0}$ & \textbf{Ours} \\
    \midrule
    $E_{Total}$ & 0.8616 & 0.9563 & 0.0006 & 0.0001 & \textbf{1.0000} \\
    \bottomrule
  \end{tabularx}
  \label{tab:alpha-decomposition}
\end{table}

\begin{figure}[t]
  \centering
  \includegraphics[width=0.48\textwidth]{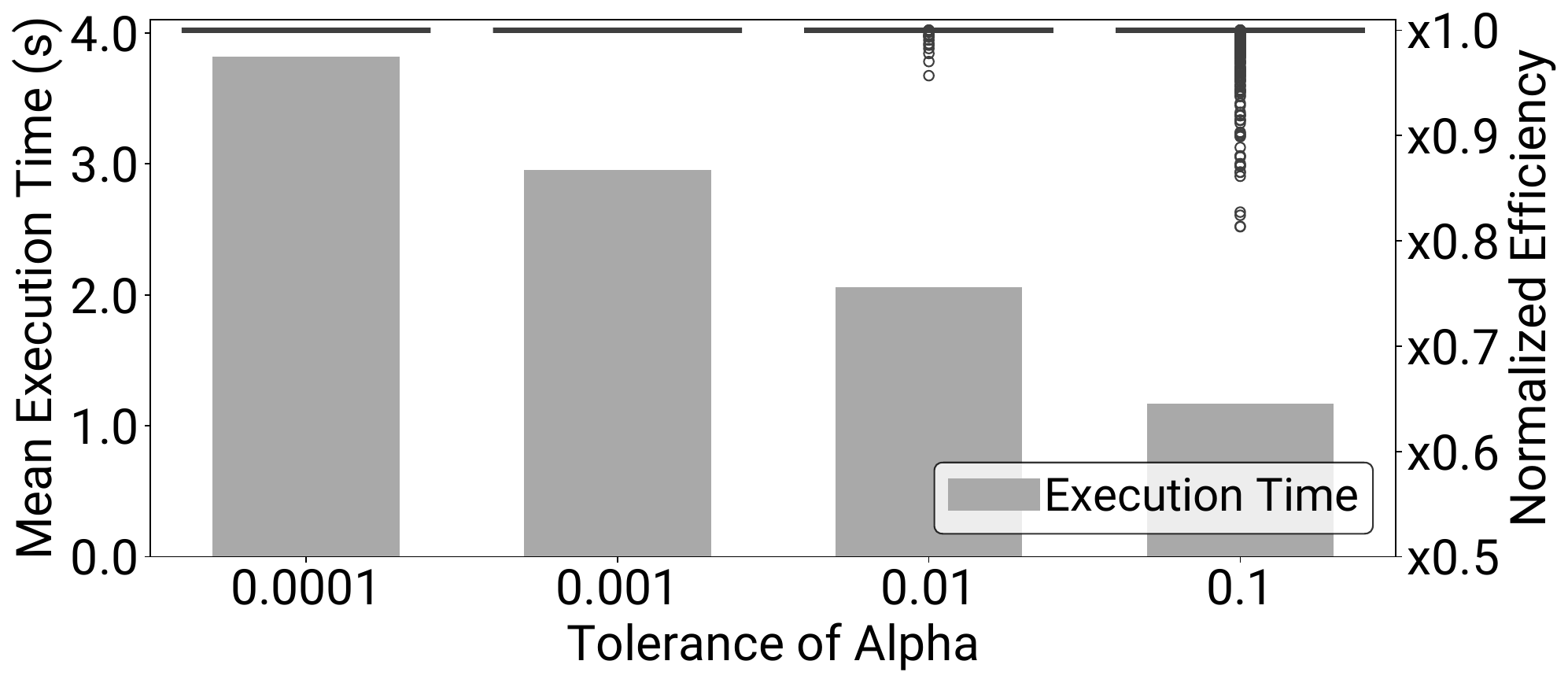}
  \caption{ILP solver latency and efficiency changes for different tolerance of $\alpha$ in the GSS algorithm}
  \Description{ILP solver latency and efficiency changes for different tolerance of $\alpha$ in the GSS algorithm}
  \label{fig:impact-of-alpha-spacing}
\end{figure}

Table~\ref{tab:alpha-decomposition} quantifies the advantage of adaptive $\alpha$ optimization by comparing normalized $E_{Total}$ across fixed $\alpha$ values and a greedy heuristic using the same input features as KubePACS. Fixed $\alpha=0.5$ and $\alpha=1.0$ suffer from severe over-provisioning, reducing $E_{Total}$ to near zero, while the greedy approach achieves only 0.86 due to a lack of global allocation control. This confirms that KubePACS's joint optimization, combining ILP-based allocation with adaptive $\alpha$ search via GSS, is essential for optimal cost-performance balance.

\paragraph{Selection of $\alpha$ Tolerance.}
Figure~\ref{fig:impact-of-alpha-spacing} analyzes the trade-off between optimal $\alpha$ search precision and ILP solver latency with respect to different tolerances, $\varepsilon$, based on 60 independent runs in the AWS N. Virginia region. As the error tolerance increases exponentially, the time to find the optimal $\alpha$ decreases linearly, which confirms the characteristics presented in Equation~\ref{eq:alpha-tolerance-range}, but at the cost of reduced $E_{Total}$. We empirically found that a tolerance of $0.01$ yields a good balance, reducing optimization time to approximately 2.0 seconds with negligible loss in recommendation quality.

\paragraph{System Overhead of the ILP Solver}
The overhead of the ILP solver was profiled over 100 iterations per region. Peak memory consumption remains under 194 MB, and average CPU utilization is limited to 1.55\%, confirming that KubePACS introduces negligible overhead to the Kubernetes provisioning pipeline.

\begin{figure}[!t]
  \centering
  \includegraphics[width=0.47\textwidth]{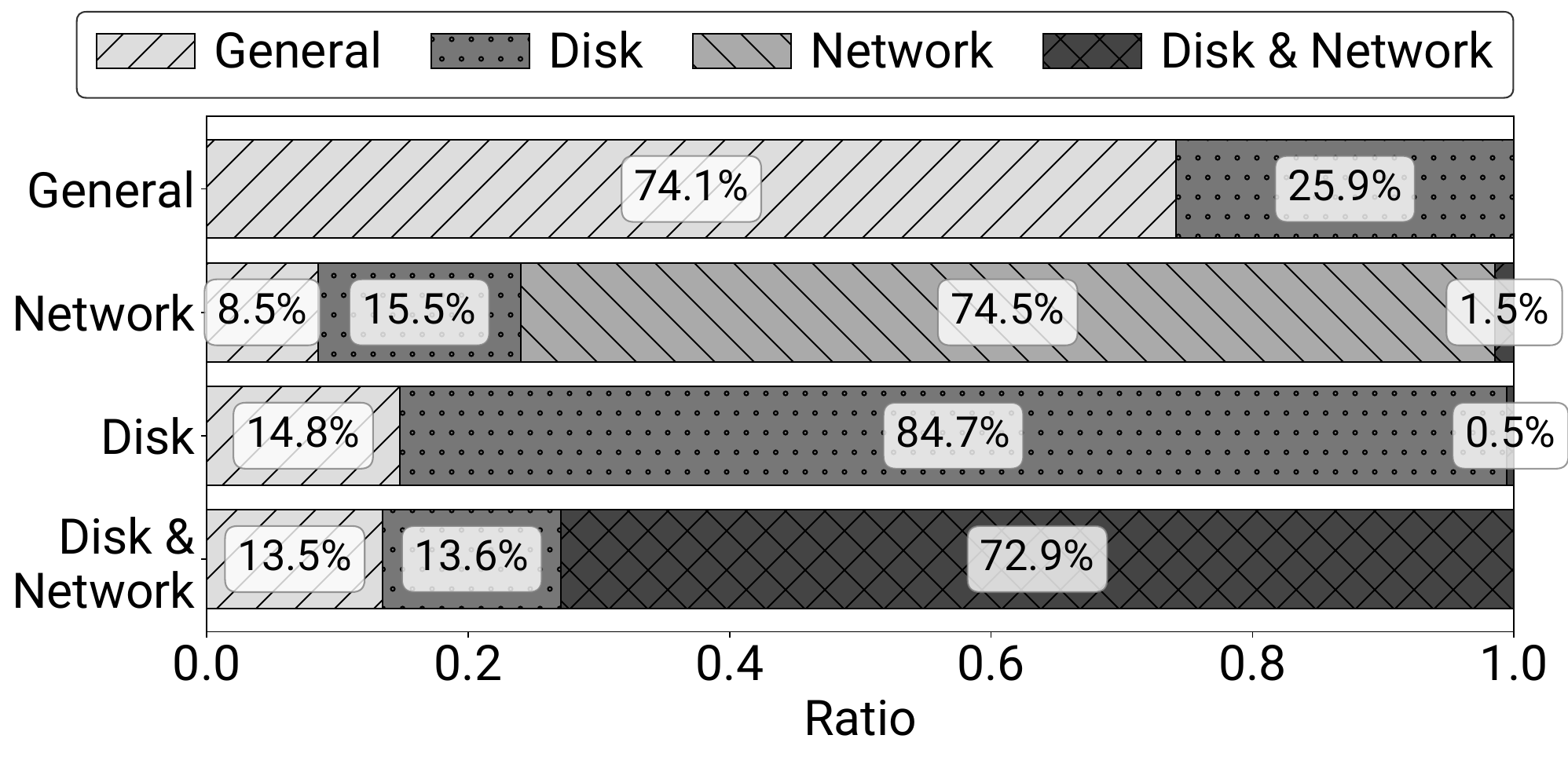}
  \caption{Effectiveness of special feature instance selection}
  \Description{Effectiveness of special feature instance selection}
  \label{fig:network-disk-intensive-workload}
\end{figure}

\paragraph{Effectiveness of Disk and Network Preferences.}
Figure~\ref{fig:network-disk-intensive-workload} illustrates the distribution of instance types selected across different network or disk I/O preference settings. In the \emph{General} scenario, where no specific preference is applied, general-purpose instances constitute the majority at 74.1\%. However, disk-optimized instances still account for 25.9\% of the selection; this is attributed to the system's cost-optimization logic, which opportunistically selects specialized instances when their spot prices drop below those of general-purpose instances. When a \emph{Network} preference is specified, the system effectively selects network-optimized instances comprising 74.5\% of the allocated nodes. Similarly, the \emph{Disk} and \emph{Disk \& Network} scenarios demonstrate strong alignment with user intent, achieving 84.7\% and 72.9\% adherence to the respective specialized instance types. These results validate the efficiency of the proposed performance scaling mechanism, demonstrating its ability to accurately map user preferences to appropriate hardware configurations while maintaining cost efficiency.

\begin{figure}[!t]
  \centering
  \includegraphics[width=0.45\textwidth]{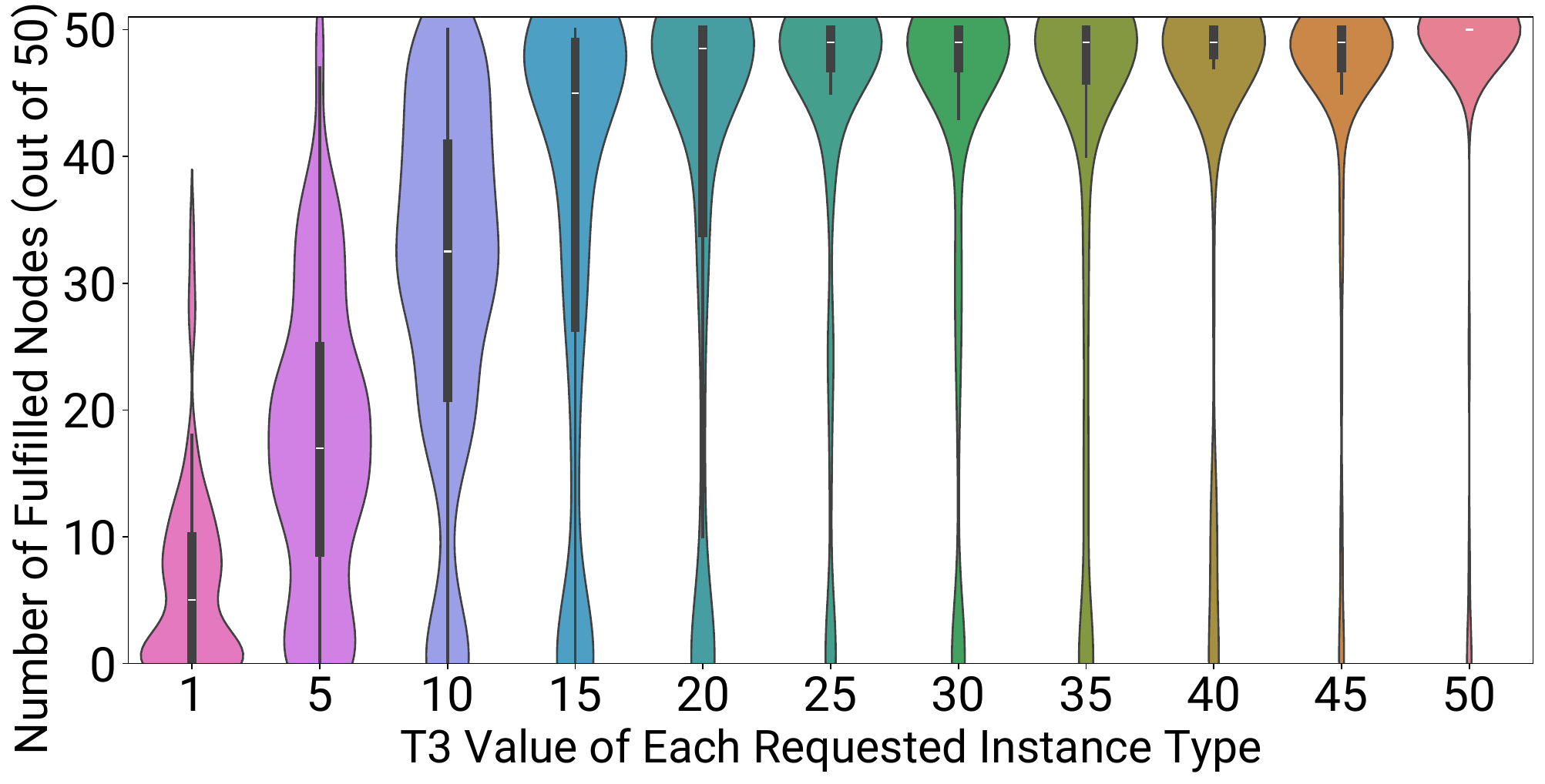}
  \caption{The number of fulfilled instances for different $T3$. The higher $T3$ ensures higher spot instance availability}
  \label{fig:t3-values-to-successful-requests-count}
\end{figure}

\paragraph{Effectiveness of Multi-node SPS Metric.}
To validate the effectiveness of the multi-node SPS metric, experiments were conducted by requesting 50 instances across varying $T3$ values at hourly intervals over a 24-hour period (May 20--21, 2025). Figure~\ref{fig:t3-values-to-successful-requests-count} illustrates the correlation between $T3$ values and the number of successfully fulfilled nodes. The results exhibit a distinct positive trend, where higher $T3$ values correspond to significantly improved success rates in spot instance provisioning. These findings justify the integration of $T3$ as a reliability constraint within the ILP formulation. By enforcing these constraints, the proposed approach ensures enhanced stability for multi-node provisioning, offering fine-grained control over single-node-centric strategies, such as SpotVerse~\cite{spotverse}.

\begin{figure*}[t]
  \centering
  \subfloat[Total hourly cost]{
    \includegraphics[height=0.1599\textwidth]{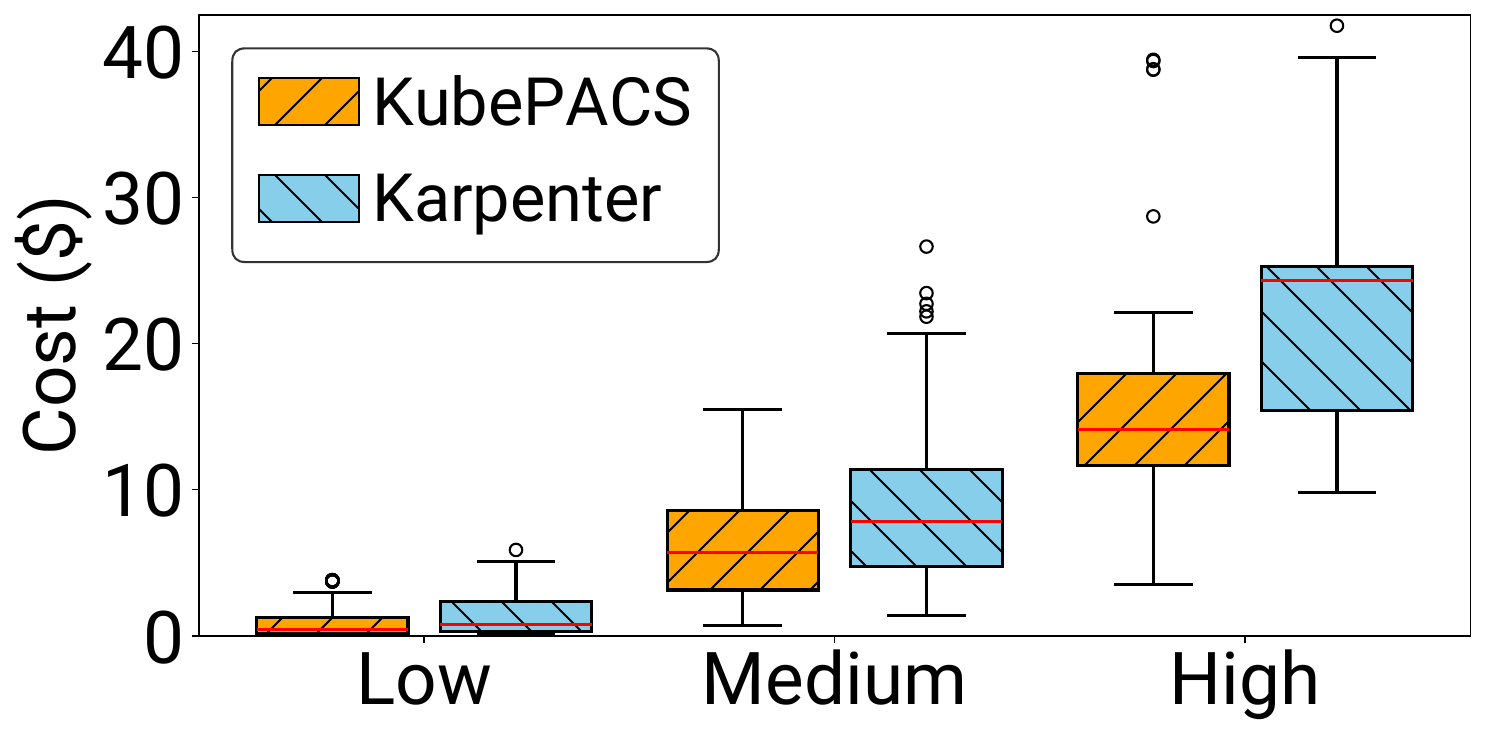}
    \label{fig:kubepacs-karpenter-cost}
  }
  \subfloat[Recommended instances benchmark score]{
    \includegraphics[height=0.1599\textwidth]{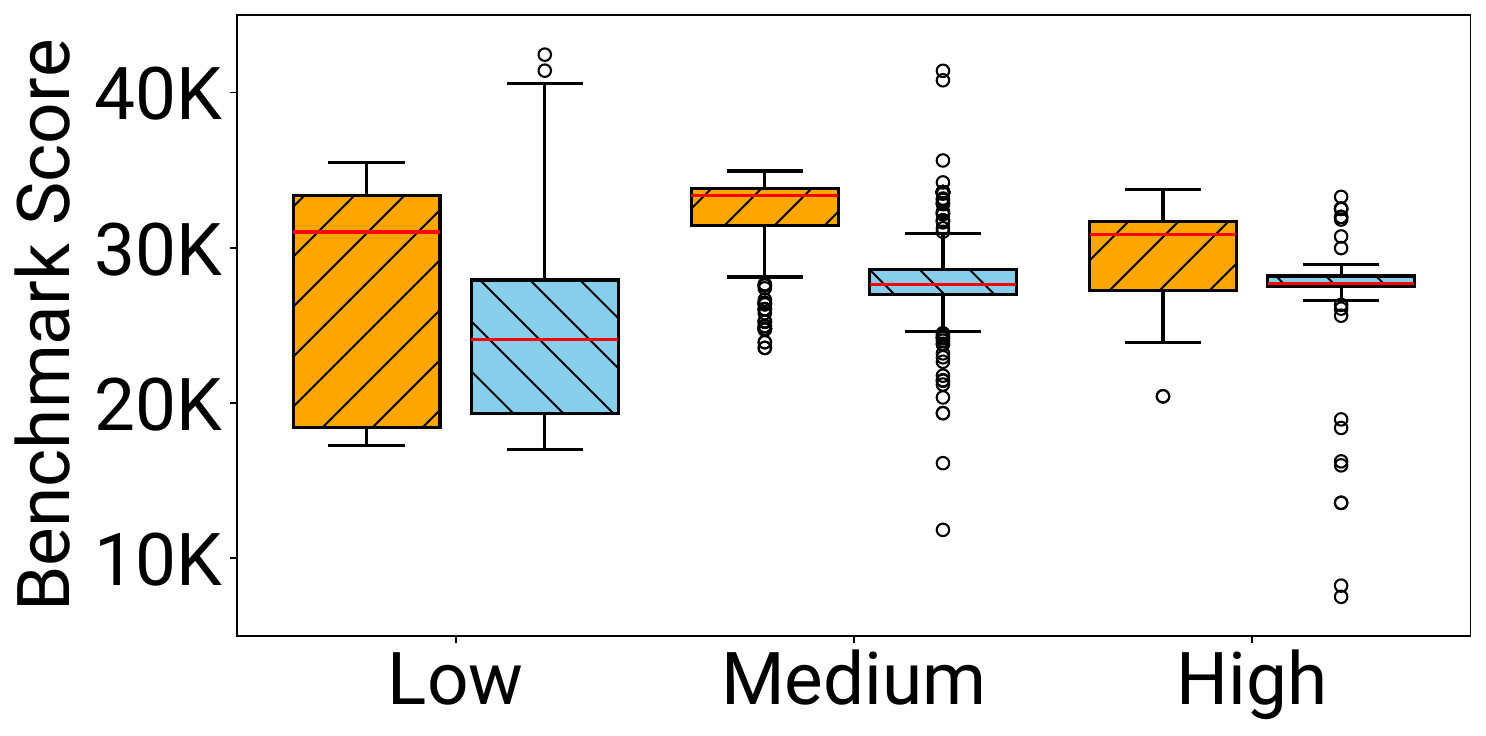}
    \label{fig:kubepacs-karpenter-performance}
  }
  \subfloat[Spot instance availability-related metrics]{
    \includegraphics[height=0.1599\textwidth]{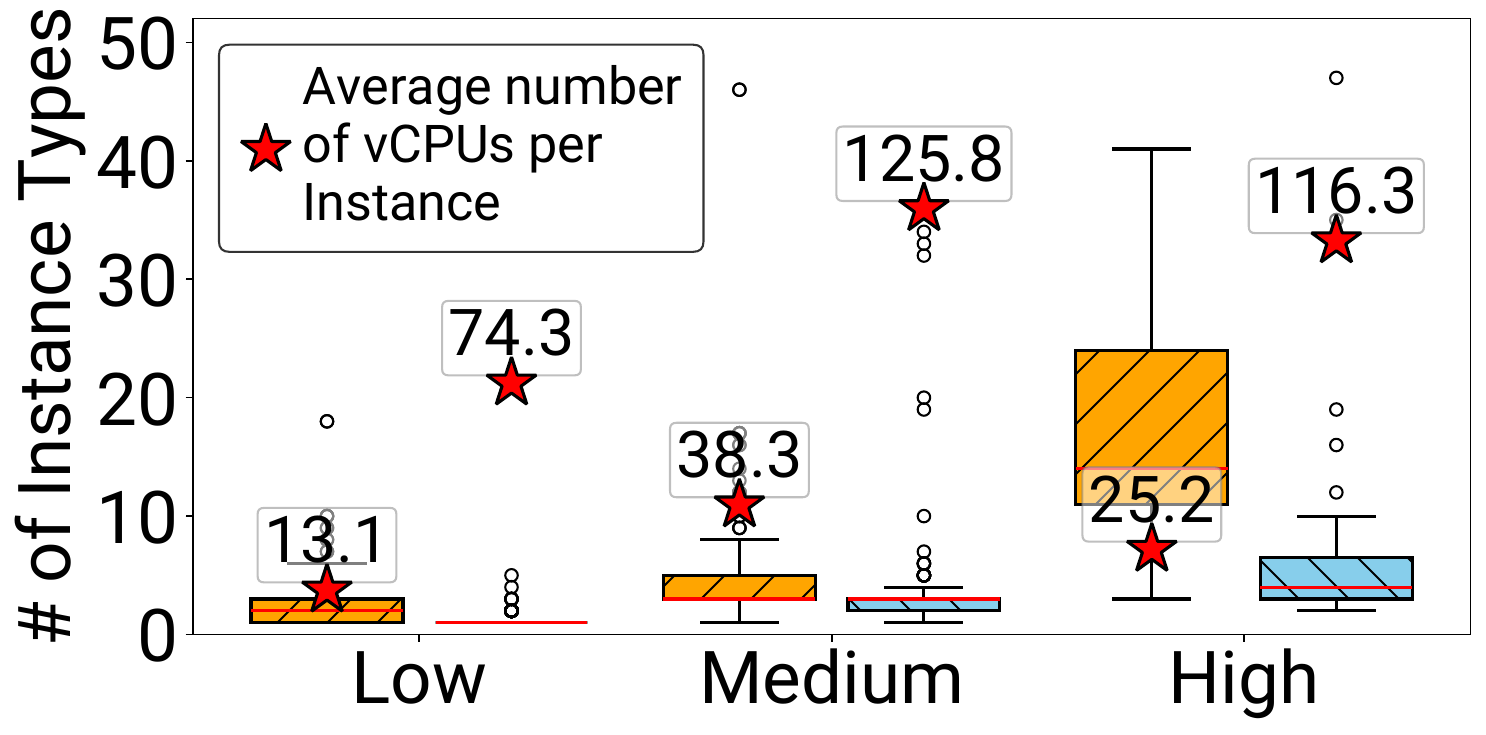}
    \label{fig:kubepacs-karpenter-availability}
  }
  \caption{{Comparison of Cost, Performance, and Availability between KubePACS and Karpenter}}
  \label{fig:kubepacs-karpenter-compare}
\end{figure*}

\subsection{KubePACS in Real-World Environments}
This section analyzes the performance of KubePACS within a realistic Kubernetes operation environment to address \textbf{RQ-3}.

\subsubsection{Comparing KubePACS with Karpenter Scheduler}
\mbox{}\\
To validate the practical applicability of the system, we conducted a comparative evaluation against Karpenter~\cite{karpenter}, a production-grade cloud instance provisioning system natively integrated with Kubernetes~\cite{kubernetes}. All experiments were conducted within AWS using clusters subject to varying pod resource demands. Workloads were categorized into three distinct intensity levels based on their aggregate CPU and memory requirements: \emph{Low} ($\le$ 200 vCPUs, $\le$ 200 GiB RAM), \emph{Medium} ($\le$ 800 vCPUs, $\le$ 4000 GiB RAM), and \emph{High} (exceeding the thresholds of the Medium category). Each scenario encompassed diverse pod configurations and was evaluated across four distinct AWS regions on August 19--20 and 22--23, 2025, as well as February 22, 2026, to capture temporal variations in spot market conditions. Each provisioning decision is independently optimized against the real-time market state at the moment of invocation, as KubePACS queries current $T3$ values and spot prices at each provisioning cycle. Figure~\ref{fig:kubepacs-karpenter-compare} presents a comprehensive comparison of KubePACS and the Karpenter scheduler from the perspectives of cost-efficiency, instance performance, and spot instance availability.

\paragraph{Cost and Performance Efficiency.}
In terms of monetary cost, KubePACS consistently demonstrates superior cost-efficiency compared to Karpenter, as illustrated in Figure~\ref{fig:kubepacs-karpenter-cost}. The experimental results indicate that KubePACS achieves an average cost reduction of 33\% by effectively identifying cost-efficient spot instances.

The reduction in cost does not come at the expense of computational capability. As shown in Figure~\ref{fig:kubepacs-karpenter-performance}, the instances recommended by KubePACS exhibit higher benchmark scores, surpassing Karpenter by an average of 12.15\%. This dual advantage confirms that KubePACS successfully optimizes the trade-off between price and performance, selecting instances that are both cost-effective and high-performing.

\paragraph{Availability of Recommended Spot Instances}
To evaluate the availability characteristics of recommended spot instances, we analyze two metrics: the cardinality of unique instance types (diversity) and the average vCPU core count per node (granularity). Prior work suggests that running workloads on a narrow set of instance types increases vulnerability to correlated interruptions~\cite{spotkube}, and larger instance types exhibit lower availability than smaller ones~\cite{spotlake-iiswc}.

Figure~\ref{fig:kubepacs-karpenter-availability} compares the diversity and granularity of instances recommended in each scenario. The vertical axis displays the distribution of unique instance types via box-and-whisker plots, while red star markers denote the average vCPU count. Karpenter tends toward consolidation, selecting few large-capacity instance types to satisfy resource demands, resulting in low type diversity and a high average vCPU count. This increases interruption risk, as losing a single large node represents a substantial loss of computational resources~\cite{spotlake-iiswc}, and reliance on a homogeneous instance set exacerbates availability risks when capacity shortages arise in a specific pool. In contrast, KubePACS mitigates these risks via $T3$-based constraints, distributing workloads across a diverse set of instance types. This diversification enables seamless substitution when specific pools experience capacity fluctuations, enhancing robustness and aggregate availability in spot-based environments.

\subsubsection{KubePACS with Real-World Applications}
\mbox{}\\
To illustrate the practical benefits of KubePACS, we present three representative real-world use cases: (1) latency-sensitive compute-intensive services, (2) large-scale batch processing workloads, and (3) applications optimized for specialized network or disk I/O hardware. These scenarios demonstrate how KubePACS outperforms alternative approaches in terms of cost-efficiency, performance, and hardware-awareness.

\paragraph{Compute-Intensive Workloads.}
A prevalent use case for Kubernetes involves the deployment of long-running services, such as REST APIs~\cite{kubernetes-review}. These services typically adopt auto-scaling via HPA~\cite{kubernetes-hpa}, which dynamically allocates additional pods based on runtime metrics (e.g., CPU utilization or incoming request rate).

For CPU-bound workloads, KubePACS facilitates the provisioning of high-performance instances at optimized costs while maintaining system reliability. To evaluate its efficiency, two representative compute-intensive REST services were deployed: a video encoding server based on ffmpeg~\cite{ffmpeg} and a continuous integration build server utilizing the Rust development toolchain.

\begin{table}[!t]
  \centering
  \setlength{\tabcolsep}{4pt}
  \renewcommand{\arraystretch}{0.9}
  \caption{Improvements for compute-intensive workloads}
  \begin{tabularx}{\linewidth}{
      >{\footnotesize\arraybackslash}c
      >{\footnotesize\arraybackslash}c
      >{\footnotesize\arraybackslash}c
      >{\footnotesize\arraybackslash}c
      >{\footnotesize\arraybackslash}c
      >{\footnotesize\arraybackslash}c
    }
    \toprule
    & \textbf{Instance} & \textbf{Price/hour} & \textbf{App} & \textbf{Req./min} & \textbf{Price/Req.} \\
    \midrule
    \multirow{2}{*}{\textbf{Karpenter}} & \multirow{2}{*}{c5.xlarge} & \multirow{2}{*}{\$0.0662} & Compilation & 9 & \$0.0074 \\
    \cmidrule(r){4-6}
    &  &  & Video enc. & 31 & \$0.0021 \\
    \midrule
    \multirow{2}{*}{\textbf{KubePACS}} & \multirow{2}{*}{c7i.xlarge} & \multirow{2}{*}{\$0.0733} & Compilation & 13 & \$0.0056 \\
    \cmidrule(r){4-6}
    &  &  & Video enc. & 47 & \$0.0016 \\
    \midrule
    \textbf{Best Case} &  & \textbf{+10.73\%} & Video enc. & \textbf{+51.61\%} & \textbf{-23.8\%} \\
    \bottomrule
  \end{tabularx}
  \label{tab:compute-services}
\end{table}

Table~\ref{tab:compute-services} presents a comparative analysis between KubePACS and Karpenter using a fixed pod configuration of 4 vCPUs \& 8 GiB RAM. A one-pod-per-instance strategy was employed to strictly isolate performance effects at the instance level. The results indicate that KubePACS selects instances with significantly higher throughput, improving request processing rates by up to 51.61\% (in the video encoding scenario), while incurring a marginal cost increase of 10.73\%. Consequently, this yields an effective performance-per-dollar gain of up to 23.8\% compared to Karpenter, while maintaining equivalent availability guarantees across both setups.

In scenarios where demand exceeds the capacity of existing pods, auto-scaling mechanisms dynamically provision additional resources. When configured to utilize KubePACS, the auto-scaler forms a fleet that inherits these aggregated benefits, ensuring a cluster that is more performant, cost-efficient, and highly available.

\paragraph{Batch Compute-Intensive Workloads.}
To assess KubePACS in batch-oriented environments, a graph analytics pipeline was implemented using the Lithops framework~\cite{lithops-tcc}, processing large-scale graph datasets from object storage via compute-intensive algorithms including Community Detection, PageRank~\cite{pagerank}, and Dijkstra~\cite{dijkstra}.

\begin{figure}[t]
  \centering
  \includegraphics[trim=0 12pt 0 0, clip, width=0.47\textwidth]{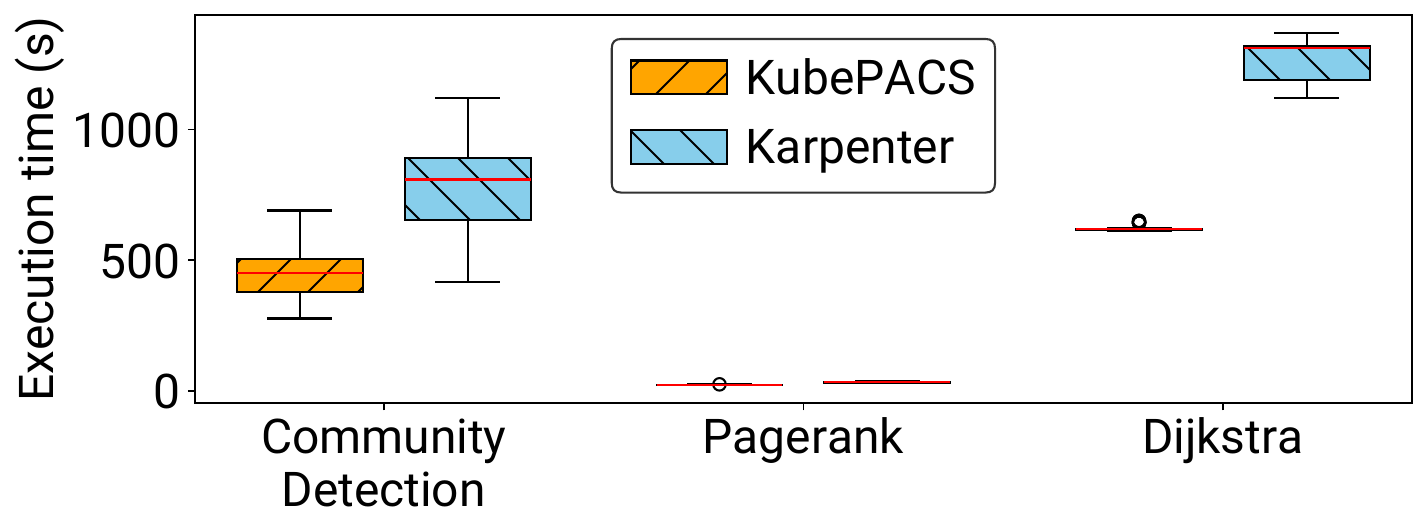}
  \caption{Execution times of graph analysis application}
  \label{fig:graph-analytics}
\end{figure}

Figure~\ref{fig:graph-analytics} illustrates the distribution of execution times for each graph algorithm under distinct instance recommendation strategies. In the baseline configuration, Karpenter, adhering to the AWS SpotFleet price-capacity-optimized policy, allocated 12 \emph{r4.xlarge} instances. Conversely, KubePACS identified and provisioned 12 \emph{r6a.\allowbreak xlarge} instances, leveraging their superior computational throughput within comparable budget constraints. As a result, the cluster managed by KubePACS demonstrates significantly reduced execution latencies across all tested workloads. This performance advantage translates to a substantial improvement in cost-effectiveness, achieving gains of up to 51.55\% compared to the Karpenter baseline.

\begin{figure}[t]
  \centering
  \subfloat[Cost comparison]{%
    \includegraphics[trim=0 10pt 0 0, clip, height=0.16\textwidth]{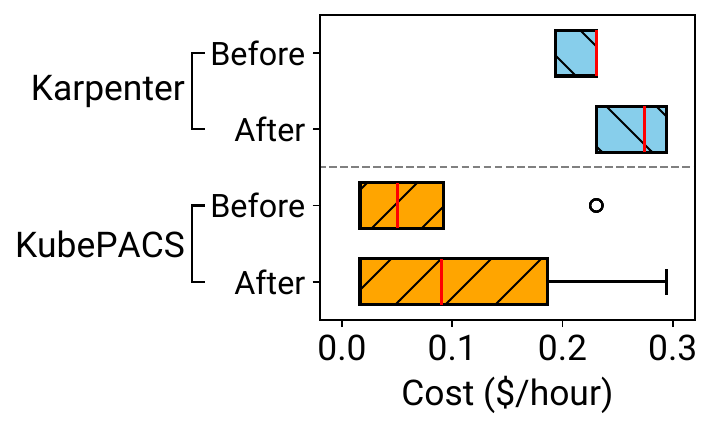}%
    \label{fig:cost-comparison}%
  }
  \hfill
  \subfloat[Performance comparison]{%
    \includegraphics[trim=0 10pt 0 0, clip, height=0.16\textwidth]{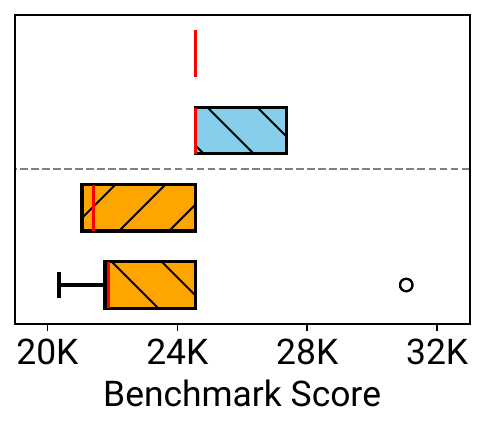}%
    \label{fig:coremark-comparison}%
  }
  \\
  \subfloat[Recovery time comparison when interruption happens]{%
    \includegraphics[trim=0 10pt 0 0, clip, width=\columnwidth]{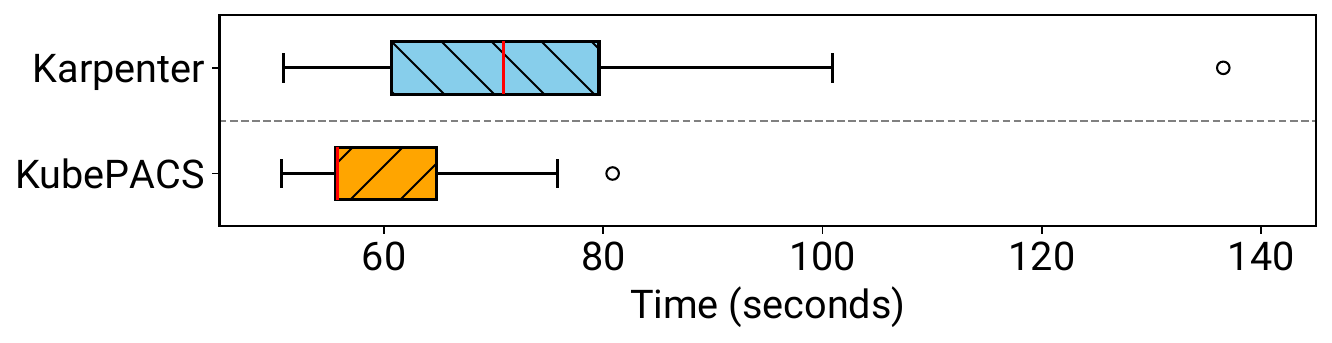}%
    \label{fig:recovery-comparison}%
  }
  \caption{The effectiveness of KubePACS interrupt handling}
  \label{fig:fis-comparisons}
\end{figure}

\paragraph{I/O-Intensive Workloads.}
KubePACS effectively accommodates I/O-intensive applications by identifying and selecting instances optimized for network or disk performance, thereby enabling both higher throughput and cost efficiency in non-CPU-intensive scenarios. Network-intensive application performance is evaluated using an Extract-Transform-Load (ETL) pipeline that ingests data from Amazon S3, an object storage service. During a parallel download of 100 GB across 100 pods, KubePACS provisions \texttt{n}-type instances, achieving a $3.16\times$ speedup compared to Karpenter, which relies on a generic provisioning policy lacking I/O awareness. This performance gain translates to an overall cost reduction of 52.44\% for the data ingestion process.

For disk-intensive tasks, the system is evaluated using a file compression workload utilizing \texttt{tar} and \texttt{gzip}. KubePACS provisions \texttt{d}-type instances characterized by superior local disk I/O throughput, resulting in $4.92\times$ faster execution relative to Karpenter, yielding a 79.86\% reduction in cost. These findings validate the efficacy of the workload-aware instance selection mechanism employed by KubePACS across diverse I/O profiles.

While Karpenter allows users to manually specify preferred instance types with specialized capabilities (e.g., high-throughput networking or NVMe SSDs) to emulate the behavior of KubePACS, identifying and maintaining an exhaustive list of such types across a vast and evolving instance catalog is operationally prohibitive. In contrast, KubePACS enables users to declare workload intents at job submission time, automatically mapping these preferences to the optimal infrastructure. The development of autonomous workload profiling mechanisms that eliminate the need for explicit user inputs remains an important direction for future research.

\subsubsection{Handling Spot Instance Interruptions}
\mbox{}\\
Figure~\ref{fig:fis-comparisons} demonstrates the effectiveness of KubePACS's interrupt handling compared to Karpenter, where interruption events were manually injected using AWS \emph{Fault Injection Service}. As shown in Figure~\ref{fig:cost-comparison}, KubePACS recommends significantly more cost-effective instances than Karpenter following an interruption event. Although Figure~\ref{fig:coremark-comparison} indicates slightly lower hardware performance, this minor tradeoff is acceptable given the substantial cost savings. Furthermore, KubePACS recovers faster than Karpenter (Figure~\ref{fig:recovery-comparison}), as Karpenter incurs considerable latency from calling the SpotFleet service for recommendations, whereas KubePACS's solver overhead is negligible.

\section{Related Work}
\textbf{Compute cluster scheduling.} Previous research has widely explored methods for cluster construction, scheduling, and workload organization, even without explicitly focusing on Kubernetes environments. Tetris~\cite{tetris} and Synergy~\cite{synergy} both assume pre-configured clusters and focus on efficiently allocating existing resources according to workload characteristics. In contrast, our work aims to construct workload-aware clusters from scratch in a cloud environment, achieving a balance between cost and performance. Stratus~\cite{stratus} and ExoSphere~\cite{exosphere} select cost-effective spot instances primarily based on task resource demands, such as CPU and memory size, but do not consider other performance-related factors. Our study proposes a method for constructing a more efficient cluster by considering not only cost and resource requirements, but also characteristics such as per-core performance, network I/O, and disk I/O. Eva~\cite{eva} dynamically optimizes the size and composition of a cloud-based cluster based on workload characteristics to achieve cost efficiency. It formulates the scheduling problem as an ILP, but solves it in practice using a reservation price-based heuristic that accounts for interference and migration overhead. However, it relies solely on on-demand instances, missing the additional cost savings that spot instances can provide.

\noindent\textbf{Enhancing spot instance usage.} Existing approaches typically optimize either cost or reliability. HotSpot~\cite{hotspot} and Proteus~\cite{proteus} aim to reduce cost via dynamic migration or hybrid scheduling, but overlook performance variability. Tributary~\cite{tributary-atc18}, Stratus~\cite{stratus}, and Can’t Be Late~\cite{spot-instance-interrupt-modeling-cant-be-late} enhance reliability through diversification or interruption modeling, yet disregard cost-performance trade-offs. SpotVerse~\cite{spotverse} uses static thresholds on SPS and IF, which do not generalize to multi-node settings. SpotKube~\cite{spotkube} jointly optimizes cost and reliability via a genetic algorithm, but lacks performance awareness and has only been tested at small scale. In contrast, this work proposes a multi-objective optimization framework that jointly considers cost, performance, and reliability. Unlike prior studies, it explicitly addresses the multi-node nature of cluster environments, utilizing real-world availability and benchmark data for instance selection, thereby improving both scalability and practical applicability.

Table~\ref{tab:novelty-comparison} summarizes the key differences between KubePACS and existing spot instance provisioning systems. Data-wise, KubePACS uniquely incorporates multi-node SPS data and hardware benchmark scores alongside spot prices. Optimization-wise, while SpotKube employs a genetic algorithm and SpotVerse applies static threshold filtering, KubePACS formulates the problem as an ILP with adaptive hyperparameter tuning via GSS and workload-aware score scaling. Systems-wise, Karpenter provides native Kubernetes integration but lacks multi-objective optimization, whereas KubePACS embeds within the Karpenter controller to enable optimized spot instance provisioning with built-in spot interruption handling.

\begin{table}[t]
  \centering
  \setlength{\tabcolsep}{2.5pt}
  \footnotesize
  \caption{{Comparison of spot instance provisioning systems}}
  \begin{tabular}{lcccc}
    \toprule
    & \textbf{Karpenter} & \textbf{SpotKube} & \textbf{SpotVerse} & \textbf{Ours} \\
    \midrule
    \textbf{Spot Price}            & \checkmark & \checkmark & \checkmark & \checkmark \\
    \textbf{SPS Awareness}       & --         & --         & Single-node & Multi-node \\
    \textbf{Benchmark Score}       & --         & --         & --         & \checkmark \\
    \textbf{Multi-obj. Optimization}     & --         & Genetic Algo. & Threshold  & ILP+GSS    \\
    \textbf{Kubernetes Integration}    & \checkmark & --         & --         & \checkmark \\
    \bottomrule
  \end{tabular}
  \label{tab:novelty-comparison}
\end{table}

\section{Discussion and Future Work}
\label{sec:future-work}
Several promising directions exist for extending the capabilities of KubePACS. While the current implementation relies on explicit user intents for instance selection, the integration of autonomous workload profiling mechanisms represents a key direction for future research. A substantial body of literature addresses workload performance profiling and prediction~\cite{mpec-cloud, cloudprophet, profet, profet-hw-metric}. Building on these foundations, we envision a two-phase approach: offline profiling that builds workload signatures from the resource utilization metrics of completed jobs, and online inference integrated into the Karpenter controller's provisioning loop to automatically classify incoming workloads. Incorporating such well-studied techniques would eliminate the need for manual specification by users, thereby enhancing usability and ensuring optimal resource mapping without human intervention.

The multi-objective optimization formulation can be extended to incorporate environmental sustainability metrics. Recent work has demonstrated carbon-aware datacenter operation at production scale~\cite{carbon-aware-gcp}, holistic carbon accounting across operational and embodied emissions~\cite{carbon-explorer}, and lifecycle-wide environmental footprint measurement~\cite{chasing-carbon}. Building on these foundations, integrating carbon intensity data into the ILP constraints would allow KubePACS to balance cost, performance, and environmental impact while preferring lower-carbon regions and instance types, fostering green computing practices without changes to the optimization structure.

Since reducing the overall number of provisioned nodes directly lowers both operational energy consumption and embodied carbon, achieving such carbon reduction goals also motivates advancing the scheduler toward cluster-level joint optimization. Future iterations will address more complex scheduling scenarios, including support for heterogeneous pod sizes and dynamic instance adjustment. Specifically, research will focus on vertical pod autoscaling mechanisms that adaptively resize allocations in response to fluctuating demand, as well as minimizing resource fragmentation when scheduling pods with diverse resource requirements on a shared node pool. By co-optimizing diverse workload types within a unified scheduler, overall node count can be reduced and resource efficiency improved, directly contributing to the carbon-aware scheduling objectives described above.

Furthermore, while the current implementation primarily targets AWS, extending KubePACS to support additional cloud providers such as Azure and GCP is a natural direction. The cross-provider experiments on Azure demonstrate that the ILP formulation generalizes effectively when analogous availability and pricing inputs are supplied, preserving the optimization behavior observed on AWS. Broadening this multi-cloud support would enable cross-provider orchestration, unlocking greater instance diversity and further opportunities for cost, performance, and carbon optimization.

Finally, extending KubePACS to support hardware accelerators, such as GPUs and TPUs~\cite{tpu}, remains an important direction. Given the high volatility and cost of accelerator-based spot instances~\cite{multinode-spot-dataset}, adapting the availability scoring metric to account for accelerator-specific instance shortages and extending the scope of orchestration across multiple regions or clouds could yield substantial benefits for large-scale AI/ML workloads.

\section{Conclusion}
In this paper, we presented KubePACS, a Kubernetes-native spot instance provisioning engine designed to optimize the selection of instance types and the number of nodes by jointly considering cost-efficiency, hardware performance, and large-scale availability. To the authors' best knowledge, this is the first work to consider a wide range of aspects of cloud spot instances to guarantee high recommendation quality. By leveraging real-time multi-node-aware datasets including SPS, spot price, and benchmark metrics, KubePACS formulates the node selection process as a multi-objective optimization problem. The system efficiently searches for an optimal trade-off using a GSS algorithm and solves the instance recommendation problem via an ILP-based approach.

We implemented KubePACS as a Helm chart using a forked version of Karpenter and evaluated it through extensive experiments with both synthetic workloads and real-world applications. Compared to state-of-the-art methods, including SpotVerse~\cite{spotverse}, SpotKube~\cite{spotkube}, and Karpenter~\cite{karpenter}, KubePACS consistently delivered higher performance per dollar, improved availability, and more efficient resource utilization. Our results demonstrate that intelligent, dataset-driven provisioning can unlock the full potential of spot instances in production Kubernetes clusters.

\begin{acks}
We sincerely thank the anonymous reviewers and our shepherd, Rüdiger Kapitza, for their invaluable feedback. We used Anthropic's Claude Code~\cite{anthropic_claude_code} to assist with the system implementation of this work. All AI-generated code was reviewed and verified by the authors. This work was supported by the National Research Foundation of Korea (NRF) grants funded by the Korea government (NRF-2020R1A2C1102544, RS-2023-00265538), the Institute of Information \& Communications Technology Planning \& Evaluation (IITP) grants funded by the Korea government (MSIT) (RS-2022-00144309, RS-2025-25441560, RS-2026-25507506, RS-2026-25548925), AI Seoul Tech Research Support Program of the Seoul Future Foundation, AWS Cloud Credits, the European Union through the Horizon Europe projects NEARDATA (101092644), CLOUDSTARS (101086248), and EXTRACT (101093110), and the Spanish Ministry of Science, Innovation and Universities through the X-AI project (PID2023-148202OB-C21).
\end{acks}

\bibliographystyle{ACM-Reference-Format}
\bibliography{kubepacs-2026}

@MISC{hadoop,
  author = {Apache Software Foundation},
  title = {Apache Hadoop},
  year = {2004},
  url  = {http://hadoop.apache.org/}
}

@article{bin-packing-algorithm,
author = {Lee, C. C. and Lee, D. T.},
title = {A Simple On-Line Bin-Packing Algorithm},
year = {1985},
issue_date = {July 1985},
publisher = {Association for Computing Machinery},
address = {New York, NY, USA},
volume = {32},
number = {3},
issn = {0004-5411},
url = {https://doi.org/10.1145/3828.3833},
doi = {10.1145/3828.3833},
journal = {J. ACM},
month = {jul},
pages = {562–572},
numpages = {11}
}

@inproceedings{spot-instance-for-hpc,
author = {Marathe, Aniruddha and Harris, Rachel and Lowenthal, David and de Supinski, Bronis R. and Rountree, Barry and Schulz, Martin},
title = {Exploiting Redundancy for Cost-Effective, Time-Constrained Execution of HPC Applications on Amazon EC2},
year = {2014},
isbn = {9781450327497},
publisher = {Association for Computing Machinery},
address = {New York, NY, USA},
url = {https://doi.org/10.1145/2600212.2600226},
doi = {10.1145/2600212.2600226},
booktitle = {Proceedings of the 23rd International Symposium on High-Performance Parallel and Distributed Computing},
pages = {279–290},
numpages = {12},
location = {Vancouver, BC, Canada},
series = {HPDC '14}
}

@inproceedings{draft-spot-instance-guarantee-from-spot-price,
author = {Wolski, Rich and Brevik, John and Chard, Ryan and Chard, Kyle},
title = {Probabilistic Guarantees of Execution Duration for Amazon Spot Instances},
year = {2017},
isbn = {9781450351140},
publisher = {Association for Computing Machinery},
address = {New York, NY, USA},
url = {https://doi.org/10.1145/3126908.3126953},
doi = {10.1145/3126908.3126953},
booktitle = {Proceedings of the International Conference for High Performance Computing, Networking, Storage and Analysis},
articleno = {18},
numpages = {11},
location = {Denver, Colorado},
series = {SC '17}
}

@article{exosphere,
  title={Portfolio-driven resource management for transient cloud servers},
  author={Sharma, Prateek and Irwin, David and Shenoy, Prashant},
  journal={Proceedings of the ACM on Measurement and Analysis of Computing Systems},
  volume={1},
  number={1},
  pages={1--23},
  year={2017},
  publisher={ACM New York, NY, USA}
}

@article{stat-analysis-spot-price,
author = {Portella, Gustavo and Rodrigues, Genaina N. and Nakano, Eduardo and Melo, Alba C.M.A.},
title = {Statistical analysis of Amazon EC2 cloud pricing models},
journal = {Concurrency and Computation: Practice and Experience},
volume = {31},
number = {18},
pages = {e4451},
doi = {https://doi.org/10.1002/cpe.4451},
url = {https://onlinelibrary.wiley.com/doi/abs/10.1002/cpe.4451},
note = {e4451 cpe.4451},
year = {2019}
}

@article{spot-analysis-javadi,
title = {Characterizing spot price dynamics in public cloud environments},
journal = {Future Generation Computer Systems},
volume = {29},
number = {4},
pages = {988-999},
year = {2013},
note = {Special Section: Utility and Cloud Computing},
issn = {0167-739X},
doi = {https://doi.org/10.1016/j.future.2012.06.012},
url = {https://www.sciencedirect.com/science/article/pii/S0167739X12001483},
author = {Bahman Javadi and Ruppa K. Thulasiram and Rajkumar Buyya}
}

@article{deconstructing-spot-instance,
author = {Agmon Ben-Yehuda, Orna and Ben-Yehuda, Muli and Schuster, Assaf and Tsafrir, Dan},
title = {Deconstructing Amazon EC2 Spot Instance Pricing},
year = {2013},
issue_date = {September 2013},
publisher = {Association for Computing Machinery},
address = {New York, NY, USA},
volume = {1},
number = {3},
issn = {2167-8375},
url = {https://doi.org/10.1145/2509413.2509416},
doi = {10.1145/2509413.2509416},
journal = {ACM Trans. Econ. Comput.},
month = {sep},
articleno = {16},
numpages = {20}
}

@article{box-whisker-plot,
  title={Variations of box plots},
  author={McGill, Robert and Tukey, John W and Larsen, Wayne A},
  journal={The American Statistician},
  volume={32},
  number={1},
  pages={12--16},
  year={1978},
  publisher={Taylor \& Francis Group}
}

@INPROCEEDINGS{profet-hw-metric,
  author={Hur, Yoonseo and Lee, Kyungyong},
  booktitle={2024 IEEE 24th International Symposium on Cluster, Cloud and Internet Computing (CCGrid)}, 
  title={CNN Training Latency Prediction Using Hardware Metrics on Cloud GPUs}, 
  year={2024},
  volume={},
  number={},
  pages={216-226},
  doi={10.1109/CCGrid59990.2024.00033}}

@INPROCEEDINGS {spotlake-iiswc,
author = {S. Lee and J. Hwang and K. Lee},
booktitle = {2022 IEEE International Symposium on Workload Characterization (IISWC)},
title = {SpotLake: Diverse Spot Instance Dataset Archive Service},
year = {2022},
volume = {},
issn = {},
pages = {242-255},
doi = {10.1109/IISWC55918.2022.00029},
url = {https://doi.ieeecomputersociety.org/10.1109/IISWC55918.2022.00029},
publisher = {IEEE Computer Society},
address = {Los Alamitos, CA, USA},
month = {nov}
}

@INPROCEEDINGS{profet,
  author={Lee, Sungjae and Hur, Yoonseo and Park, Subin and Lee, Kyungyong},
  booktitle={2022 IEEE International Conference on Big Data (Big Data)}, 
  title={PROFET: PROFiling-based CNN Training Latency ProphET for GPU Cloud Instances}, 
  year={2022},
  volume={},
  number={},
  pages={186-193},
  doi={10.1109/BigData55660.2022.10020212}}

@ARTICLE{lithops-tcc,
  author={Sampé, Josep and Sánchez-Artigas, Marc and Vernik, Gil and Yehekzel, Ido and García-López, Pedro},
  journal={IEEE Transactions on Cloud Computing}, 
  title={Outsourcing Data Processing Jobs With Lithops}, 
  year={2023},
  volume={11},
  number={1},
  pages={1026-1037},
  doi={10.1109/TCC.2021.3129000}
}

@inproceedings{alibaba-spot-instance,
author = {Movsowitz Davidow, Danielle and Agmon Ben-Yehuda, Orna and Dunkelman, Orr},
title = {Deconstructing Alibaba Cloud's Preemptible Instance Pricing},
year = {2023},
isbn = {9798400701559},
publisher = {Association for Computing Machinery},
address = {New York, NY, USA},
url = {https://dl.acm.org/doi/pdf/10.1145/3588195.3593001},
booktitle = {Proceedings of the 32nd International Symposium on High-Performance Parallel and Distributed Computing},
pages = {253–265},
numpages = {13},
location = {Orlando, FL, USA},
series = {HPDC '23}
}

@inproceedings{multinode-spot-dataset,
  author       = {Sungkyu Cheon and Kyumin Kim and Kyunghwan Kim and Moohyun Song and Kyungyong Lee},
  title        = {Multi-Node Spot Instances Availability Score Collection System},
  booktitle    = {Proceedings of the 34th International Symposium on High-Performance Parallel and Distributed Computing (HPDC '25)},
  year         = {2025},
  publisher    = {ACM},
  month        = {July}, 
}

@inproceedings{multi-spotlake-www,
author = {Kim, Kyunghwan and Park, Subin and Hwang, Jaeil and Lee, Hyeonyoung and Kang, Seokhyeon and Lee, Kyungyong},
title = {Public Spot Instance Dataset Archive Service},
year = {2023},
isbn = {9781450394192},
publisher = {Association for Computing Machinery},
address = {New York, NY, USA},
url = {https://doi.org/10.1145/3543873.3587314},
doi = {10.1145/3543873.3587314},
booktitle = {Companion Proceedings of the ACM Web Conference 2023},
pages = {69–72},
numpages = {4},
location = {Austin, TX, USA},
series = {WWW '23 Companion}
}

@inproceedings{single-node-sps-spot-interruption-visible,
author = {Kim, KyungHwan and Lee, Kyungyong},
title = {Making Cloud Spot Instance Interruption Events Visible},
year = {2024},
isbn = {9798400701719},
publisher = {Association for Computing Machinery},
address = {New York, NY, USA},
url = {https://doi.org/10.1145/3589334.3645548},
doi = {10.1145/3589334.3645548},
booktitle = {Proceedings of the ACM Web Conference 2024},
pages = {2998–3009},
numpages = {12},
location = {Singapore, Singapore},
series = {WWW '24}
}

@inproceedings {spot-instance-interrupt-modeling-cant-be-late,
author = {Zhanghao Wu and Wei-Lin Chiang and Ziming Mao and Zongheng Yang and Eric Friedman and Scott Shenker and Ion Stoica},
title = {Can{\textquoteright}t Be Late: Optimizing Spot Instance Savings under Deadlines},
booktitle = {21st USENIX Symposium on Networked Systems Design and Implementation (NSDI 24)},
year = {2024},
isbn = {978-1-939133-39-7},
address = {Santa Clara, CA},
pages = {185--203},
url = {https://www.usenix.org/conference/nsdi24/presentation/wu-zhanghao},
publisher = {USENIX Association},
month = apr
}

@inproceedings{spot-price-change-2017,
author = {Baughman, Matt and Caton, Simon and Haas, Christian and Chard, Ryan and Wolski, Rich and Foster, Ian and Chard, Kyle},
title = {Deconstructing the 2017 Changes to AWS Spot Market Pricing},
year = {2019},
isbn = {9781450367585},
publisher = {Association for Computing Machinery},
address = {New York, NY, USA},
url = {https://doi.org/10.1145/3322795.3331465},
doi = {10.1145/3322795.3331465},
booktitle = {Proceedings of the 10th Workshop on Scientific Cloud Computing},
pages = {19–26},
numpages = {8},
location = {Phoenix, AZ, USA},
series = {ScienceCloud '19}
}

@INPROCEEDINGS{spot-price-policy-change-2017-irwin,  author={Irwin, David and Shenoy, Prashant and Ambati, Pradeep and Sharma, Prateek and Shastri, Supreeth and Ali-Eldin, Ahmed},  booktitle={2019 28th International Conference on Computer Communication and Networks (ICCCN)},   title={The Price Is (Not) Right: Reflections on Pricing for Transient Cloud Servers},   year={2019},  volume={},  number={},  pages={1-9},  doi={10.1109/ICCCN.2019.8846933}}

@misc{geekbench,
  author = {Primate Labs},
  title = {Geekbench 6 - Cross-Platform Benchmark},
  howpublished = "\url{https://www.geekbench.com/}",
  year = {2025}
}

@misc{spec-benchmark,
  author = {Standard Performance Evaluation Corporation},
  title = {SPEC Benchmarks and Tools},
  howpublished = "\url{https://www.spec.org/benchmarks.html}",
  year = {2023}
}

@misc{coremark,
  author = {EEMBC},
  title = {CPU Benchmark – MCU Benchmark – CoreMark – EEMBC Embedded Microprocessor Benchmark Consortium},
  howpublished = "\url{https://www.eembc.org/coremark/}",
  year = {2025}
}

@misc{azure-coremark,
  author = {Microsoft Azure},
  title = {Compute benchmark scores for Azure Linux VMs},
  howpublished = "\url{https://learn.microsoft.com/en-us/azure/virtual-machines/linux/compute-benchmark-scores}",
  year = {2025}
}

@misc{gcp-coremark,
  author = {Google Cloud Platform},
  title = {CoreMark scores of VM instances by family},
  howpublished = "\url{https://cloud.google.com/compute/docs/coremark-scores-of-vm-instances}",
  year = {2025}
}

@INPROCEEDINGS{mpec-cloud,
author = {M. Son and K. Lee},
booktitle = {2018 IEEE 11th International Conference on Cloud Computing (CLOUD)},
title = {Distributed Matrix Multiplication Performance Estimator for Machine Learning Jobs in Cloud Computing},
year = {2018},
volume = {00},
number = {},
pages = {638-645},
doi = {10.1109/CLOUD.2018.00088},
url = {doi.ieeecomputersociety.org/10.1109/CLOUD.2018.00088},
ISSN = {2159-6190},
month={Jul}
}

@book{kubernetes,
    title	= {Kubernetes - Scheduling the Future at Cloud Scale},
    author	= {David K. Rensin},
    year	= {2015},
    URL	= {http://www.oreilly.com/webops-perf/free/kubernetes.csp},
    booktitle	= {OSCON 2015},
    pages	= {All},
    address	= {1005 Gravenstein Highway North Sebastopol, CA 95472}
}

@inproceedings{pywren,
     author = {Jonas, Eric and Pu, Qifan and Venkataraman, Shivaram and Stoica, Ion and Recht, Benjamin},
     title = {Occupy the Cloud: Distributed Computing for the 99\%},
     booktitle = {Proceedings of the 2017 Symposium on Cloud Computing},
     series = {SoCC '17},
     year = {2017},
     isbn = {978-1-4503-5028-0},
     location = {Santa Clara, California},
     pages = {445--451},
     numpages = {7},
     url = {http://doi.acm.org/10.1145/3127479.3128601},
     doi = {10.1145/3127479.3128601},
     acmid = {3128601},
     publisher = {ACM},
     address = {New York, NY, USA},
}

@inproceedings{cloud-usage-pattern,
author = {Kilcioglu, Cinar and Rao, Justin M. and Kannan, Aadharsh and McAfee, R. Preston},
title = {Usage Patterns and the Economics of the Public Cloud},
year = {2017},
isbn = {9781450349130},
publisher = {International World Wide Web Conferences Steering Committee},
address = {Republic and Canton of Geneva, CHE},
url = {https://doi.org/10.1145/3038912.3052707},
doi = {10.1145/3038912.3052707},
booktitle = {Proceedings of the 26th International Conference on World Wide Web},
pages = {83–91},
numpages = {9},
location = {Perth, Australia},
series = {WWW '17}
}

@techreport{pagerank,
          number = {1999-66},
           month = {November},
          author = {Lawrence Page and Sergey Brin and Rajeev Motwani and Terry Winograd},
            note = {Previous number = SIDL-WP-1999-0120},
           title = {The PageRank Citation Ranking: Bringing Order to the Web.},
            type = {Technical Report},
       publisher = {Stanford InfoLab},
            year = {1999},
     institution = {Stanford InfoLab},
             url = {http://ilpubs.stanford.edu:8090/422/}
}

@INPROCEEDINGS{spot-price-by-location,  author={Ekwe-Ekwe, Nnamdi and Barker, Adam},  booktitle={2018 18th IEEE/ACM International Symposium on Cluster, Cloud and Grid Computing (CCGRID)},   title={Location, Location, Location: Exploring Amazon EC2 Spot Instance Pricing Across Geographical Regions},   year={2018},  volume={},  number={},  pages={370-373},  doi={10.1109/CCGRID.2018.00059}}

@inproceedings{spot-instance-analysis,
author = {Wang, Cheng and Liang, Qianlin and Urgaonkar, Bhuvan},
title = {An Empirical Analysis of Amazon EC2 Spot Instance Features Affecting Cost-Effective Resource Procurement},
year = {2017},
isbn = {9781450344043},
publisher = {Association for Computing Machinery},
address = {New York, NY, USA},
url = {https://doi.org/10.1145/3030207.3030210},
doi = {10.1145/3030207.3030210},
booktitle = {Proceedings of the 8th ACM/SPEC on International Conference on Performance Engineering},
pages = {63–74},
numpages = {12},
location = {L'Aquila, Italy},
series = {ICPE '17}
}

@article{tpu,
author = {Jouppi, Norman P. and Young, Cliff and Patil, Nishant and Patterson, David and others},
title = {In-Datacenter Performance Analysis of a Tensor Processing Unit},
year = {2017},
issue_date = {May 2017},
publisher = {Association for Computing Machinery},
address = {New York, NY, USA},
volume = {45},
number = {2},
issn = {0163-5964},
url = {https://doi.org/10.1145/3140659.3080246},
doi = {10.1145/3140659.3080246},
journal = {SIGARCH Comput. Archit. News},
month = jun,
pages = {1–12},
numpages = {12}
}

@article{normalize,
  author       = {S. Gopal Krishna Patro and
                  Kishore Kumar Sahu},
  title        = {Normalization: {A} Preprocessing Stage},
  journal      = {CoRR},
  volume       = {abs/1503.06462},
  year         = {2015},
  url          = {http://arxiv.org/abs/1503.06462},
  eprinttype    = {arXiv},
  eprint       = {1503.06462},
  bibsource    = {dblp computer science bibliography, https://dblp.org}
}

@MISC{spot-placement-score-start,
  author = {AWS What is New},
  title = {Introducing Amazon EC2 Spot placement score},
  year = {2021},
  url  = {https://aws.amazon.com/about-aws/whats-new/2021/10/amazon-ec2-spot-placement-score/}
}

@MISC{azure-spot-placement-score,
  author = {Azure},
  title = {Spot Placement Score},
  year = {2025},
  url  = {https://learn.microsoft.com/en-us/azure/virtual-machine-scale-sets/spot-placement-score}
}

@inproceedings {tributary-atc18,
author = {Aaron Harlap and Andrew Chung and Alexey Tumanov and Gregory R. Ganger and Phillip B. Gibbons},
title = {Tributary: spot-dancing for elastic services with latency {SLOs}},
booktitle = {2018 USENIX Annual Technical Conference (USENIX ATC 18)},
year = {2018},
isbn = {978-1-931971-44-7},
address = {Boston, MA},
pages = {1--14},
url = {https://www.usenix.org/conference/atc18/presentation/harlap},
publisher = {USENIX Association},
month = jul
}

@MISC{spotfleet,
  author = {AWS},
  title = {EC2 Fleet and Spot Fleet},
  year = {2024},
  url  = {https://docs.aws.amazon.com/AWSEC2/latest/UserGuide/Fleets.html}
}

@inproceedings{spotverse,
author = {Son, Myungjun and Akbulut, Gulsum Gudukbay and Kandemir, Mahmut Taylan},
title = {SpotVerse: Optimizing Bioinformatics Workflows with Multi-Region Spot Instances in Galaxy and Beyond},
year = {2024},
isbn = {9798400706233},
publisher = {Association for Computing Machinery},
address = {New York, NY, USA},
url = {https://doi.org/10.1145/3652892.3700750},
doi = {10.1145/3652892.3700750},
booktitle = {Proceedings of the 25th International Middleware Conference},
pages = {74–87},
numpages = {14},
location = {Hong Kong, Hong Kong},
series = {Middleware '24}
}

@INPROCEEDINGS{spotkube,
  author={Edirisinghe, Dasith and Rajapakse, Kavinda and Abeysinghe, Pasindu and Rathnayake, Sunimal},
  booktitle={2024 IEEE International Conference on Cloud Computing Technology and Science (CloudCom)}, 
  title={SpotKube: Cost-Optimal Microservices Deployment with Cluster Autoscaling and Spot Pricing}, 
  year={2024},
  volume={},
  number={},
  pages={87-94},
  doi={10.1109/CloudCom62794.2024.00026}
}

@article{ffmpeg,
author = {Tomar, Suramya},
title = {Converting video formats with FFmpeg},
year = {2006},
issue_date = {June 2006},
publisher = {Belltown Media},
address = {Houston, TX},
volume = {2006},
number = {146},
issn = {1075-3583},
journal = {Linux J.},
month = jun,
pages = {10}
}

@article{dijkstra,
  title={A note on two problems in connexion with graphs},
  author={Dijkstra, Edsger W},
  journal={Numerische mathematik},
  volume={1},
  number={1},
  pages={269--271},
  year={1959},
  publisher={Springer}
}

@MISC{karpenter,
  author = {Karpenter community},
  title = {Karpenter : Just-in-time Nodes for Any Kubernetes Cluster},
  year = {2025},
  url  = {https://karpenter.sh/}
}

@book{golden-section-search,
author = {Press, William H. and Teukolsky, Saul A. and Vetterling, William T. and Flannery, Brian P.},
title = {Numerical Recipes 3rd Edition: The Art of Scientific Computing. Chapter 10.1. Golden Section Search in One Dimension},
year = {2007},
isbn = {0521880688},
publisher = {Cambridge University Press},
address = {USA},
edition = {3},
}

@inproceedings{kubernetes-hpa,
author = {Baresi, Luciano and Hu, Davide Yi Xian and Quattrocchi, Giovanni and Terracciano, Luca},
title = {KOSMOS: Vertical and Horizontal Resource Autoscaling for Kubernetes},
year = {2021},
isbn = {978-3-030-91430-1},
publisher = {Springer-Verlag},
address = {Berlin, Heidelberg},
url = {https://doi.org/10.1007/978-3-030-91431-8_59},
doi = {10.1007/978-3-030-91431-8_59},
booktitle = {Service-Oriented Computing: 19th International Conference, ICSOC 2021, Virtual Event, November 22–25, 2021, Proceedings},
pages = {821–829},
numpages = {9},
location = {Dubai, United Arab Emirates}
}

@inproceedings{eva,
author = {Chang, Tzu-Tao and Venkataraman, Shivaram},
title = {Eva: Cost-Efficient Cloud-Based Cluster Scheduling},
year = {2025},
isbn = {9798400711961},
publisher = {Association for Computing Machinery},
address = {New York, NY, USA},
url = {https://doi.org/10.1145/3689031.3717483},
doi = {10.1145/3689031.3717483},
booktitle = {Proceedings of the Twentieth European Conference on Computer Systems},
pages = {1399–1416},
numpages = {18},
location = {Rotterdam, Netherlands},
series = {EuroSys '25}
}

@inproceedings{hotspot,
author = {Shastri, Supreeth and Irwin, David},
title = {HotSpot: automated server hopping in cloud spot markets},
year = {2017},
isbn = {9781450350280},
publisher = {Association for Computing Machinery},
address = {New York, NY, USA},
url = {https://doi.org/10.1145/3127479.3132017},
doi = {10.1145/3127479.3132017},
booktitle = {Proceedings of the 2017 Symposium on Cloud Computing},
pages = {493–505},
numpages = {13},
location = {Santa Clara, California},
series = {SoCC '17}
}

@inproceedings{proteus,
author = {Harlap, Aaron and Tumanov, Alexey and Chung, Andrew and Ganger, Gregory R. and Gibbons, Phillip B.},
title = {Proteus: agile ML elasticity through tiered reliability in dynamic resource markets},
year = {2017},
isbn = {9781450349383},
publisher = {Association for Computing Machinery},
address = {New York, NY, USA},
url = {https://doi.org/10.1145/3064176.3064182},
doi = {10.1145/3064176.3064182},
booktitle = {Proceedings of the Twelfth European Conference on Computer Systems},
pages = {589–604},
numpages = {16},
location = {Belgrade, Serbia},
series = {EuroSys '17}
}

@inproceedings{stratus,
author = {Chung, Andrew and Park, Jun Woo and Ganger, Gregory R.},
title = {Stratus: cost-aware container scheduling in the public cloud},
year = {2018},
isbn = {9781450360111},
publisher = {Association for Computing Machinery},
address = {New York, NY, USA},
url = {https://doi.org/10.1145/3267809.3267819},
doi = {10.1145/3267809.3267819},
booktitle = {Proceedings of the ACM Symposium on Cloud Computing},
pages = {121–134},
numpages = {14},
location = {Carlsbad, CA, USA},
series = {SoCC '18}
}

@article{golden-section-method,
author = {Gupta, Murli},
year = {1991},
month = {03},
pages = {},
title = {Numerical Methods and Software (David Kahaner, Cleve Moler, and Stephen Nash)},
volume = {33},
journal = {Siam Review - SIAM REV},
doi = {10.1137/1033033}
}

@INPROCEEDINGS{golden-section-equation,
  author={Chang, Yen-Ching},
  booktitle={2009 2nd International Conference on Biomedical Engineering and Informatics}, 
  title={N-Dimension Golden Section Search: Its Variants and Limitations}, 
  year={2009},
  volume={},
  number={},
  pages={1-6},
  doi={10.1109/BMEI.2009.5304779}
}

@misc{pulp,
  author = {Stuart A. Mitchell},
  title = {PuLP},
  howpublished = "\url{https://github.com/coin-or/pulp}",
  year = {2003},
}

@book{genetic-algorithm,
  address = {New York},
  author = {Goldberg, David E.},
  publisher = {Addison-Wesley},
  title = {Genetic Algorithms in Search, Optimization, and Machine Learning},
  username = {dalbem},
  year = 1989
}

@misc{kubernetes-review,
      title={Benefits, Challenges, and Research Topics: A Multi-vocal Literature Review of Kubernetes}, 
      author={Shazibul Islam Shamim and Jonathan Alexander Gibson and Patrick Morrison and Akond Rahman},
      year={2022},
      eprint={2211.07032},
      archivePrefix={arXiv},
      primaryClass={cs.SE},
      url={https://arxiv.org/abs/2211.07032}, 
}

@inproceedings{tetris,
author = {Grandl, Robert and Ananthanarayanan, Ganesh and Kandula, Srikanth and Rao, Sriram and Akella, Aditya},
title = {Multi-resource packing for cluster schedulers},
year = {2014},
isbn = {9781450328364},
publisher = {Association for Computing Machinery},
address = {New York, NY, USA},
url = {https://doi.org/10.1145/2619239.2626334},
doi = {10.1145/2619239.2626334},
booktitle = {Proceedings of the 2014 ACM Conference on SIGCOMM},
pages = {455–466},
numpages = {12},
location = {Chicago, Illinois, USA},
series = {SIGCOMM '14}
}

@inproceedings {synergy,
author = {Jayashree Mohan and Amar Phanishayee and Janardhan Kulkarni and Vijay Chidambaram},
title = {Looking Beyond {GPUs} for {DNN} Scheduling on {Multi-Tenant} Clusters},
booktitle = {16th USENIX Symposium on Operating Systems Design and Implementation (OSDI 22)},
year = {2022},
isbn = {978-1-939133-28-1},
address = {Carlsbad, CA},
pages = {579--596},
url = {https://www.usenix.org/conference/osdi22/presentation/mohan},
publisher = {USENIX Association},
month = jul
}

@ARTICLE{cloudprophet,
  author={Huang, Darong and Costero, Luis and Pahlevan, Ali and Zapater, Marina and Atienza, David},
  journal={IEEE Transactions on Sustainable Computing}, 
  title={CloudProphet: A Machine Learning-Based Performance Prediction for Public Clouds}, 
  year={2024},
  volume={9},
  number={4},
  pages={661-676},
  doi={10.1109/TSUSC.2024.3359325}}

@article{carbon-aware-gcp,
  title={Carbon-aware computing for datacenters},
  author={Radovanovi{\'c}, Ana and Koningstein, Ross and Schneider, Ian and Chen, Bokan and Duarte, Alexandre and Roy, Binz and Xiao, Diyue and Haridasan, Maya and Hung, Patrick and Care, Nick and others},
  journal={IEEE Transactions on Power Systems},
  volume={38},
  number={2},
  pages={1270--1280},
  year={2022},
  publisher={IEEE}
}

@inproceedings{carbon-explorer,
author = {Acun, Bilge and Lee, Benjamin and Kazhamiaka, Fiodar and Maeng, Kiwan and Gupta, Udit and Chakkaravarthy, Manoj and Brooks, David and Wu, Carole-Jean},
title = {Carbon Explorer: A Holistic Framework for Designing Carbon Aware Datacenters},
year = {2023},
isbn = {9781450399166},
publisher = {Association for Computing Machinery},
address = {New York, NY, USA},
url = {https://doi.org/10.1145/3575693.3575754},
doi = {10.1145/3575693.3575754},
booktitle = {Proceedings of the 28th ACM International Conference on Architectural Support for Programming Languages and Operating Systems, Volume 2},
pages = {118–132},
numpages = {15},
location = {Vancouver, BC, Canada},
series = {ASPLOS 2023}
}

@article{chasing-carbon,
author = {Gupta, Udit and Kim, Young Geun and Lee, Sylvia and Tse, Jordan and Lee, Hsien-Hsin S. and Wei, Gu-Yeon and Brooks, David and Wu, Carole-Jean},
title = {Chasing Carbon: The Elusive Environmental Footprint of Computing},
year = {2022},
issue_date = {July-Aug. 2022},
publisher = {IEEE Computer Society Press},
address = {Washington, DC, USA},
volume = {42},
number = {4},
issn = {0272-1732},
url = {https://doi.org/10.1109/MM.2022.3163226},
doi = {10.1109/MM.2022.3163226},
journal = {IEEE Micro},
month = jul,
pages = {37–47},
numpages = {11}
}

@misc{anthropic_claude_code,
  author       = {{Anthropic}},
  title        = {Claude Code},
  year         = {2026},
  howpublished = {\url{https://www.anthropic.com/claude-code}},
}

\end{document}